\documentclass[twocolumn,tighten]{aastex61}
\usepackage{amsmath,amssymb,amstext}
\usepackage[all]{hypcap} 
\usepackage{amssymb,amsmath,amsthm}
\usepackage{longtable}
\usepackage{amssymb}
\usepackage{amsmath, xspace}
\usepackage{graphics,graphicx} 
\usepackage{rotating}
\usepackage{color}
\usepackage{multirow}

\def\CfA{Center for Astrophysics $|$ Harvard \& Smithsonian,
60 Garden Street, Cambridge, MA 02138}
\def\MIT{Kavli Institute for Astrophysics and Space
Research, Massachusetts Institute of Technology, 77 Massachusetts Avenue,
Cambridge, MA 02139}
\def\MPE{Max Planck Institute for Extraterrestrial Physics, Giessenbachstr. 1 85748, Garching, Germany}
\def\INAF{INAF/IASF-Milano, Via Bassini 15, 20133 Milano, Italy}
\def\Penn{Pennsylvania State University, Department of Astronomy and Astrophysics, 525 Davey Lab, University Park, Pennsylvania 16802, USA}
\def\SU{Department of Physics, Stanford University, 382 Via Pueblo Mall, Stanford, CA 94305-4060, USA}
\def\icrar{ICRAR, University of Western Australia, 35 Stirling Hwy, Crawley, WA 6009, Australia}
\def\erlangen{Dr. Remeis-Sternwarte and Erlangen Centre for Astroparticle Physics, Universit\"at Erlangen-Nürnberg, Sternwartstr. 7, 96049 Bamberg, Germany}
\def\openu{Centre for Electronic Imaging, The Open University, Walton Hall, Milton Keynes, MK7 6AA, UK}
\def\INAFB{INAF Osservatorio di Astrofisica e Scienza dello Spazio di Bologna, via Gobetti 93/3, 40129 Bologna, Italy}

\renewcommand{\sc}{\footnotesize} 
\newcommand{\athena}{{\it Athena}}
\newcommand{\xmm}{{\it XMM-Newton}}

\newcommand{\chandra}{{\it Chandra}}
\newcommand{\erosita}{{\it eROSITA}}
\begin{document}

\title{Characterisation of the Particle-Induced Background of XMM-Newton EPIC-pn: \\Short and Long Term Variability}

\author{Esra~Bulbul}
\affiliation{\CfA}
\author{Ralph~Kraft}
\affiliation{\CfA}
\author{Paul~Nulsen}
\affiliation{\CfA}
\affiliation{\icrar}
\author{Michael~Freyberg}
\affiliation{\MPE}
\author{Eric~D.~Miller}
\affiliation{\MIT}
\author{Catherine~Grant}
\affiliation{\MIT}
\author{Mark~W.~Bautz}
\affiliation{\MIT}
\author{David~N.~Burrows}
\affiliation{\Penn}
\author{Steven~Allen}
\affiliation{\SU}
\author{Tanja~Eraerds}
\affiliation{\MPE}
\author{Valentina~Fioretti}
\affiliation{\INAFB}
\author{Fabio~Gastaldello}
\affiliation{\INAF}
\author{Vittorio~Ghirardini}
\affiliation{\CfA}
\author{David~Hall}
\affiliation{\openu}
\author{Norbert~Meidinger}
\affiliation{\MPE}
\author{Silvano~Molendi}
\affiliation{\INAF}
\author{Arne~Rau}
\affiliation{\MPE}
\author{Dan~Wilkins}
\affiliation{\SU}
\author{Joern~Wilms}
\affiliation{\erlangen}

\correspondingauthor{Esra~Bulbul}
\email{ebulbul@cfa.harvard.edu}

\begin{abstract}
The particle-induced background of X-ray observatories is produced by Galactic Cosmic Ray (GCR) primary protons, electrons, and He ions. Events due to direct interaction with the detector are usually removed by on board processing. The interactions of these primary particles with the detector environment produce secondary particles that mimic X-ray events from celestial sources and are much more difficult to identify. The filter wheel closed data from the {\it XMM-Newton} EPIC-pn camera in small window mode (SWM) contains both the X-ray-like background events and the events due to direct interactions with the primary particles. From this data we demonstrate that X-ray-like background events are spatially correlated with the primary particle interaction. This result can be used to further characterise and reduce the non-X-ray background in silicon-based X-ray detectors in current and future missions. We also show that spectrum and pattern fractions of secondary particle events are different from those produced by cosmic X-rays.

\end{abstract}
\section{Introduction}

X-ray studies of the assembly processes of extended large scale structures, constraints on cosmology and the nature of dark matter, and studies of the cosmic X-ray background that holds clues about the formation of the first black holes are among the primary science goals of current (e.g., \chandra, \xmm, and SRG) and future X-ray telescopes (\athena, {\it Lynx}) \citep{nandra13, gaskin19}. These measurements are sensitive to the level of the total flux and related systematic uncertainties of the instrumental X-ray background. Understanding, accurately characterizing, and reducing the absolute level of this X-ray background are fundamental to the X-ray analysis of faint X-ray sources and deep surveys.

The X-ray background can be classified into two major components: the cosmic X-ray background (CXB) and particle-induced non-X-ray background (NXB). The CXB is dominated by three main components: the Galactic local foreground, solar wind charge exchange emission, and unresolved X-ray emission by distant celestial sources. At lower energies ($<$1~keV) the dominant component is thermal emission from the Galactic Halo contributing at intermediate and high Galactic latitudes \citep{burrows91, snowden91, warwick02, lumb02} and the Local Hot Bubble, a region of hot plasma ($T\,\sim\,10^{6}$ K) mostly filling the local cavity extending 100~pc away from the Sun \citep{snowden98}. Another component, which is composed of C~{\sc VI}, O~{\sc VII}, O ~{\sc VIII}, Ne~{\sc IX}, and Mg~{\sc XI} line emission at lower energies ($<1$~keV), is the solar wind charge exchange produced when highly charged solar wind ions interact with neutral atoms in the solar system \citep{robertson03,koutroumpa06}. Unresolved X-ray emission from distant astrophysical sources, e.g., Active Galactic Nuclei (AGNs), contributes a power-law continuum spectrum that dominates at higher energies ($>$1~keV) with a possible change in slope at lower energies and has been extensively studied in the literature, e.g., \citet{lumb02,moretti09}. The magnitude of this component varies with position on the sky and it clearly suffers from cosmic variance \citep{hickox07}. If an observation is deep enough to resolve the brightest sources (e.g., the strong shot noise), the residual contribution reaches to the expected cosmic variance given the Log $N$ -- log $S$ relation (see for example Fig. 9 in \citep{moretti09} and discussion therein). 

The non X-ray background due to particles in missions operating above the Earth's magnetic belts consists of two major background components: soft protons focused by the mirrors onto the focal plane and particle-induced instrumental background. Soft protons that are generated in the Solar corona and in the Earth's magnetosphere with energies less than a few 100~keV can follow the optical path through the telescope and be focused onto the detectors. The spectral shape of this component can be described by a power-law continuum with highly variable magnitude and slope \citep{kuntz08}. When present, soft protons can increase the  total background intensity by three orders of magnitude on short timescales of 10--10$^{4}$~s \citep{kuntz08}. They deposit most of their energy near the surface of the detector and produce valid event patterns \citep{gastaldello17}. 

On the other hand, the unfocused particle-induced internal detector background is generated by energetic Galactic Cosmic Ray (GCR) primaries with energies from several tens of MeV to several GeV. GCR particles consisting of protons, electrons, and He ions are subject to variations over the Solar cycle. These incoming particles interact with the detector and produce secondary particles. The interactions constitute the major components of the unfocused portion of the particle-induced instrumental background \citep{kuntz08, snowden08, vonkienlin18}. Based on their high total energies or the pattern of pixels excited in the event, particle events generated by primary GCRs are mostly discarded on board by the event processing (e.g., by the Minimum Ionizing Particle, MIP, rejection algorithm for \xmm) to prevent them from saturating the limited bandwidth for telemetry \citep{lumb02}. However, the secondary electrons and photons due to this unfocused component deposit charge in the detector that it is challenging to distinguish from X-ray events from celestial sources and thus contribute significantly to (and often dominate) the quiescent instrumental background. 

Quantifying the particle-induced instrumental background of X-ray observatories is not a trivial process and needs careful examination of observations while the detector is not exposed to sky. The \xmm\ observatory, carrying two types of silicon-based X-ray detectors on board, the European Photon Imaging Camera (EPIC) MOS \citep{turner01} and the EPIC pn \citep{struder01}, provides an excellent opportunity to explore the instrumental background of silicon-based X-ray detectors. The unexposed corners of the \xmm\ EPIC MOS detector that are masked off and the MOS data obtained when the filter wheel is in the closed position (FWC data) serve as estimators of the particle background for each observation that are used in the X-ray analysis of faint extended sources \citep{deluca04, kuntz08, gastaldello17}. The particle background of \xmm\ EPIC-pn is difficult to predict and eliminate due to the fact that the unexposed region on the detector is small, i.e., statistics on the background level is limited. 

In this paper, we examine the long-term variability of the unfocused EPIC-pn background. We present results from an analysis of all archival EPIC-pn data in the small window mode (SWM) with the filter wheel closed and MIP rejection disabled. The filter wheel closed observations with 1.05~mm of Al shielding do not allow any photons from celestial sources or soft protons to reach the focal plane. Additionally, all of the pixel data from both valid events and normally rejected particle tracks (GCR primaries) are telemetered to the ground in SWM mode observations. This set-up provides a unique opportunity to quantitatively investigate the relationship between the energetic primaries (i.e., GCRs) and the secondaries that mimic X-rays from celestial sources which constitute the dominant component of the instrumental background. We describe our sample and data analysis methods in Section \ref{sec:analysis}. Our results for the \xmm\ EPIC-pn SWM observations are described in Section \ref{sec:result_smw}. Our conclusions are given in Section \ref{sec:concl}.

\begin{figure*}
\centering
\includegraphics[width=0.99\textwidth]{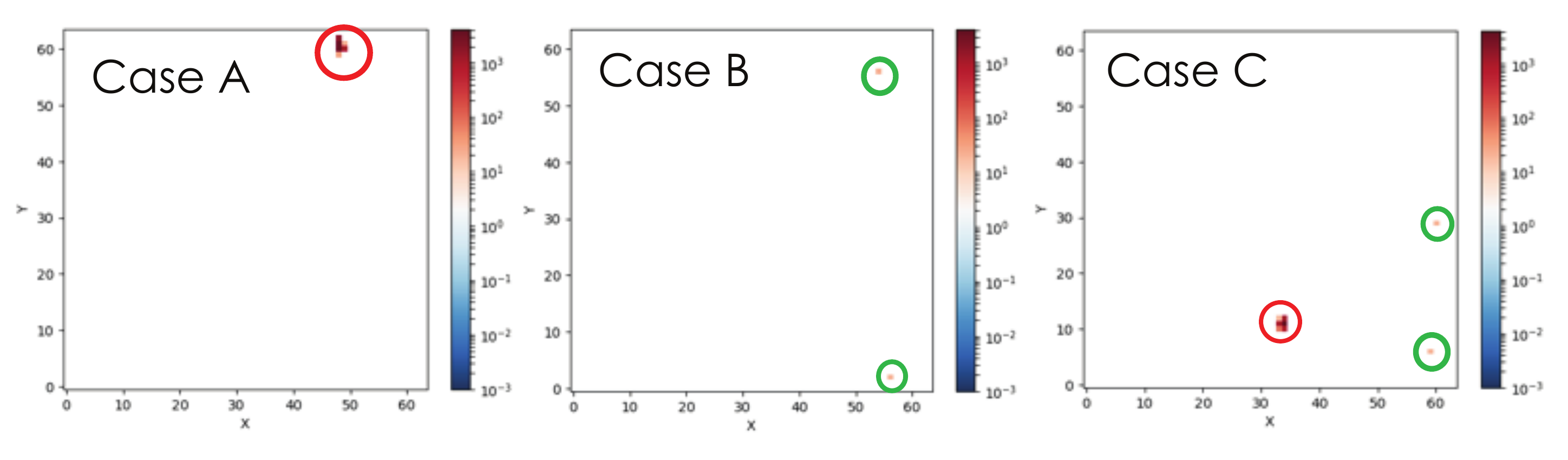}
\caption{Frames with just particle tracks (Case~A), valid events (Case~B), and both particle tracks and valid events (Case~C) are shown.
The circles in red mark the detected primary particle events, while the green circles show the secondary valid events.}
\label{fig:frames}
\vspace{4mm}
\end{figure*}

This work was originally performed as part of a program to develop algorithms for improved background characterization and reduction for the \athena\ Wide-Field Imager (WFI) Science Products Module (SPM) \citep{burrows18,bulbul18,grant18}. One of the goals of the SPM would have been to use the full data stream from the WFI, not just the ground science event data available to the observer, to reduce the instrumental background. In an effort to better understand the instrumental background in X-ray observatories, we examined the \xmm\ EPIC-pn SWM data as described in this paper and modeled the WFI background using the GEANT4 software \citep{tenzer2010}. The latter modeling was done using the measured particle background at the \athena\ orbit (L2) with a mass model of the flight instrument \citep{vonkienlin18}. Results from this study will be presented in a separate publication \citep{miller20}.

\section{XMM-Newton EPIC-pn Data Analysis}
\label{sec:analysis}
\subsection{Filter-Wheel Closed Slew and Pointed Observations}

The EPIC-pn CCD camera is one of the primary instruments on on-board \xmm, with a collecting area of $\sim$2500 cm$^{2}$ at 1~keV and a 27.2~arcmin by 26.2~arcmin field of view over the broad energy range of 0.1~keV to 12~keV \citep{struder01}. The \xmm\ EPIC-pn data used in this work were taken during slews when the filter wheel was closed and performed in the SWM. In this mode, a 63~pixel by 64~pixel (4.3~arcmin by 4.4~arcmin) region on detector CCD4 is active and  the read-out time is 5.67~ms, roughly a factor of 13 faster than the full-frame readout time of the primary science observing mode (full frame mode). A total of 309 observations have been completed since 2007 between revolutions 1360--3217, with typical exposures of 3--7~ks, adding up to a total exposure time of 1~Ms. The observation IDs and exposure times of the slew FWC observations are given in Table \ref{table:obs}.

FWC observations are performed with 1.05~mm aluminum shielding, preventing low-energy soft protons and X-rays from celestial sources from reaching the EPIC-pn detector. Thus, FWC exposures contain only the particle-induced internal detector background, generated as a result of interactions of energetic Galactic Cosmic Rays (E$>$100~MeV) with the material surrounding the EPIC-pn camera. Additionally, in the observations taken in the SWM set-up, the standard minimum ionizing particle (MIP) rejection algorithm, which identifies and automatically eliminates the pixels above a certain energy threshold and invalid patterns identified on board from the telemetered data, is inactive. As a result, these observations represent an ideal data set to characterize the long term behavior of the \xmm\ EPIC-pn internal background, since the ground observer has full access to {\it all} pixels above a threshold set by the ground observer. The data consist of electronic readout noise (at lowest energies, hot  pixels, columns, and readout noise), primary high energy GCRs, secondaries generated by high energy galactic cosmic rays, and particle-induced X-rays \citep[continuum and fluorescent lines,][]{vonkienlin18}.  

\begin{figure*}
\centering
\includegraphics[width=\textwidth]{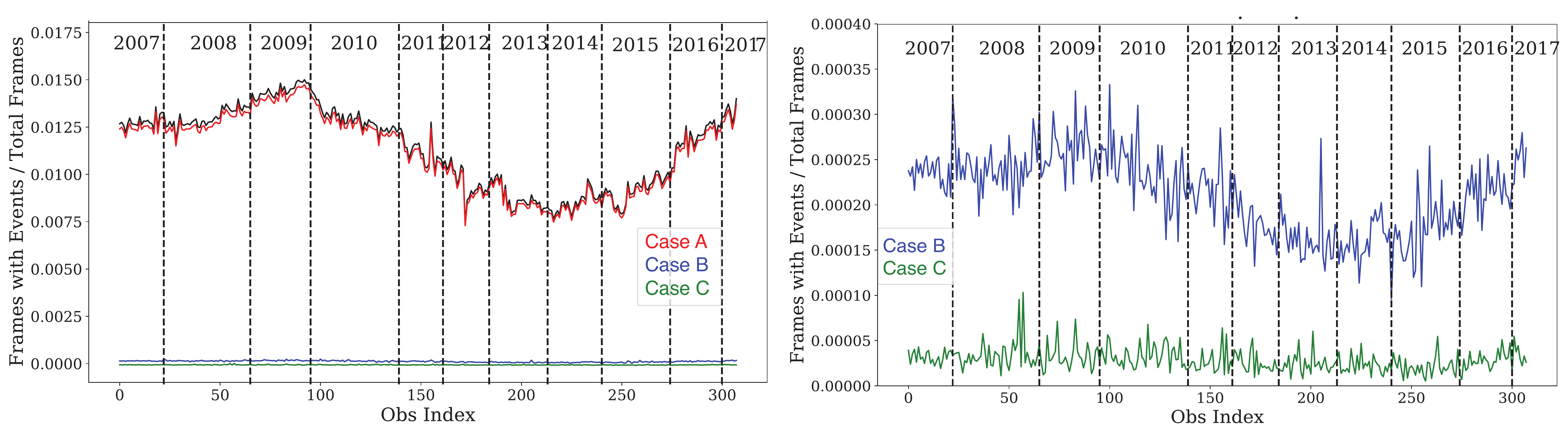}
\caption{Left: The fraction of frames with just primary particle events (red; Case~A), just secondary valid events (blue; Case~B), and with both valid and particle events (green; Case~C) as a function of time. Curve in black shows the total rate of particle events. Strong modulation with the Solar cycle observed for all the frames indicates that the \xmm\ EPIC-pn unfocused background is dominated by Galactic cosmic rays. Right: Zoom onto Case~B and Case~C frame to enhance the visibility of the Solar modulation cycle.}
\label{fig:var_frames}
\vspace{2mm}
\end{figure*}

The SWM frame time is sufficiently short in a sufficiently small readout area that the particle rate is much smaller than the frame rate. We thus have the unique opportunity to associate the normally rejected charged-particle events with the valid events that comprise the instrumental background. Since these observations are mostly dominated by the unfocused X-ray background, we use the term non-X-ray background (NXB) for these FWC slew observations hereafter. 

For a comparison, we also examine the pointed \xmm\ EPIC-pn SWM observations from two celestial sources: observations of the AB~Doradus star system with the closed and thick filters, and a supernova remnant G21.5$-$0.9 (SNR 21.5$-$09, hereafter) performed with the thin filter. The details of these observations are given in Table~\ref{table:obs_pointed}. The AB~Doradus observations with the closed filter are not expected to include any source photons (NXB dominated), while the observation with the thick filter is expected to be dominated by soft protons. A detailed analysis of the comparison between these two observations with different set-up will be explored in another article (Bulbul et al., in prep). SNR 21.5$-$09 observations, on the other hand, are dominated by photons from the supernova remnant in the FOV in the 2--7~keV band, while the contribution from the non X-ray background is subdominant. These pointed observations, taken in the SWM set-up, are similarly telemetered to the ground with the ``on-board'' MIP rejection algorithm inactive, thus including all pixels above the energy threshold. Having a longer uninterrupted  exposure time, these data provide information on short term variability of the unfocused X-ray background.

\subsection{Data Reduction and Analysis}

We first run the standard the Science Analysis Software {\it SAS} algorithm {\it epchain} to eliminate hot pixels and columns from the data and to form event lists for single exposures and for a given list of CCDs from the relevant observation data files (ODF) \citep{gabriel2004}. We note that a non-standard parameter setting is selected in the {\it epchain} runs to switch off the ``on-ground" MIP rejection.  We then construct individual frames from the event files using the frame rate of 5.67~ms. The total number of frames constructed is given in Table \ref{table:obs}. We examined a total of $\sim$1.86$\times10^8$ frames in the \xmm\ EPIC-pn slew observations in this work. These observations span 10 years covering a full solar cycle. 

After the construction of frames, we run an image segmentation algorithm on each frame to identify the independent event islands. This algorithm finds connected pixels and traces the long charge tracks of the energetic particle interactions. The charge of each event island is determined by the total charge enclosed in that particular event island. The centroids of these event islands are defined by the maximally charged pixel. We then assign a pattern ID by the pattern detection algorithm, i.e., {\it epchain}, to each event island. We note that this image segmentation algorithm developed by our team, is not the same algorithm used by the on board software.

The standard \xmm\ EPIC-pn event processing flags event islands with pattern ID$\leq$12 as valid events, while particle events are marked with pattern IDs $>12$\footnote{https://xmm-tools.cosmos.esa.int/external/xmm\_user\_support\\/documentation/uhb/epic\_evgrades.html}. The pattern ID is related to the number and pattern of the CCD pixels triggered for an X-ray event above a certain threshold. The pattern IDs with 0 mark valid single pixel events, double pixel events are marked with pattern IDs 1--4, while triple and quadruple events have pattern IDs of 5--8 and 8--12, respectively. We note that in this FWC data set, the valid events are dominated by secondary particles that are produced by interactions of primary GCRs with the surroundings of the instrument. This component is mostly composed of secondary electrons and photons that deposit their energy in the active volume of the detector. The contributing secondary electrons are generated in ionization processes, while the secondary photons are mainly generated in bremsstrahlung and inelastic scattering processes
(see \citealt{vonkienlin18} for more detail).  In this work, we only consider valid events in the 2--7~keV energy band to avoid low-energy detector noise, unless otherwise noted. 
The event islands marked with pattern IDs $>$12 are mostly the incoming background GCR particles ($\sim$200~MeV--GeV), and Supra-Thermal Ions (STIs), mostly protons, accelerated in the Heliosphere to energies up to $<$100~keV hitting the detector \citep{vonkienlin18}. These particle event islands are identified based on their patterns and the total charge encapsulated within the island. For most of these energetic events, there exists more than one pixel with a total charge exceeding the saturation level of the analog-to-digital converter (ADC; corresponding to 22.5~keV when MIP rejection is off). In those cases, the centroid of a particle event island is the maximally charged pixel last read by the image segmentation algorithm.

\subsection{Classification of Frames}
We next analyze the data sets on a frame-by-frame basis and identify event islands and divide the frames into four categories:  frames with just particle tracks (Case~A), frames with only valid events (Case~B), frames with at least one particle track and at least one valid event (Case~C), and frames with no particle tracks or events (Case D). This categorization allow us to examine the frames with particle primaries without a secondary (Case~A), secondaries that are created by particle primaries but not detected on the same frame (Case~B), and the frames with the primary particle events and their secondaries detected on the same frame (Case~C). Figure \ref{fig:frames} illustrates the subdivision of frames.
We find that overall the total number of Case~A frames is 2089948, while 39186 of the frames are in the Case~B, and 5175 of the frames are in Case~C categories.

\begin{table}
\centering
\caption{Fractions of Frames}
\label{table:frame_class}
\begin{tabular}{lcc}
\hline\hline
Frame Type & Number of Frames & $\%$ Fractions \\ 
\hline
Case~A & 2089948 & 1.12\\
Case~B & 39186 & 0.02 \\
Case~C & 5175 & 0.003 \\
Case D & 184541368 & 98.8 \\
\hline\hline
\end{tabular}
\end{table}
\section{Results}
\label{sec:result_smw}

In the next subsections, we investigate the spectral properties, light curves, and spatial correlations between valid  events and particle tracks in Case~A, B, and C frames in detail. 

\subsection{Long Term Variability}
Investigating the temporal changes in the number of frames, we find that the overwhelming majority of the frames are empty and fall under the Case D category independent of solar cycle or orbit (see Figure 2).  We find that overall the total number of Case~A frames is 2089948 (1.12\% of the total), while 39186 (0.02\%) of the frames are Case~B, 5175 (0.003\%) of the frames are Case~C, and the remainder (98.8\%) are empty Case D frames (see \ref{table:frame_class}).
The temporal changes in the fractional A, B, and C frame rates are shown in Figure \ref{fig:var_frames}. The clear modulation with solar cycle observed in the fraction of Case~A, B, and C frames is consistent with the previously observed modulations in the count rates in unexposed corners of MOS2 detector, EPIC Radiation Monitor on board of \xmm\ \citep{gastaldello17}, and \chandra\ high energy (12--15 keV) count rate for the ACIS-S3 CCD as a function of year.\footnote{http://space.mit.edu/cgrant/cti/cti120/bkg.pdf} GCR flux is modulated in anti-correlation with solar activity due to the solar wind \citep{neher62}. While, Tthe Earth's magnetic field provides a varying degree of geomagnetic shielding from these GCR particles, the level of the modulation depends on the energy of GCRs. The observed modulation on the EPIC-pn data shows that the FWC data are dominated by GCRs.

\begin{table*}
\centering
\caption{Pattern Distribution of valid events in Case~B and Case~C frames of the Non-Xray Background (NXB) taken in filter-wheel closed set up, AB~Doradus observations in filter-wheel closed set up, and SNR 21.5$-$09 observations performed with thin filter.}
\label{table:branching}
\begin{tabular}{l|ccc|ccc|cc}
\hline\hline
Frames  & NXB       &   & & AB~Doradus  & & & SNR 21.5$-$09   & \\
        & Closed Flt.&  & & Closed Flt. & & & Thin Flt.     & \\
        &  Case~B & Case~C  & &  Case~B & Case~C &&  Case~B & Case~C \\
        \hline
Singles & 65.6~$\pm$~0.2 & 67.8~$\pm$~0.8 &&  65.3~$\pm$~0.7  &  69.4~$\pm$~2.6 && 61.6~$\pm$~0.1 & 60.7~$\pm$~0.7\\ 
Doubles & 31.3~$\pm$~0.2  & 30.1~$\pm$~0.7 && 32.1~$\pm$~0.7 & 28.5~$\pm$~2.5  && 34.5~$\pm$~0.1 & 35.9~$\pm$~0.7\\
Triples & 1.47~$\pm$~0.06 & 1.0~$\pm$~0.2 && 1.0~$\pm$~0.1 & 1.0~$\pm$~0.6 && 2.0~$\pm$~0.6 & 1.7~$\pm$~0.2 \\
Quadruples & 1.59~$\pm$~0.06 & 1.1~$\pm$~0.2 && 1.6~$\pm$~0.2 & 1.0~$\pm$~0.6  && 1.9~$\pm$~0.6 & 1.6~$\pm$~0.2 \\
\hline\hline
\end{tabular}
\end{table*}
\subsection{Branching Ratios of Valid Events}
\label{sec:branchingsec}

Having the largest number of valid events, Case~B frames dominate the unfocused background of the \xmm\ EPIC-pn. We next examine the branching ratios, i.e., the fraction of the patterns of the valid events (singles, doubles, triples, and quadruples) in Case~B frames, where only secondary events are detected. We detect a total number of 39190 valid events in non X-ray background observations (i.e., slew filter-wheel-closed observations listed in Table \ref{table:obs}). The fractions of these valid patterns in Case~B frames are shown in Table \ref{table:branching}. Of the total valid events, $65.6 \pm 0.2\%$ of them are singles, $31.3 \pm 0.2\%$ are doubles, while $1.47 \pm 0.06\%$ and $1.59 \pm 0.06\%$ are triples and quadruples, respectively. Comparing these ratios with Case~B frames observed in the closed filter AB~Doradus observations, of the total 4172 valid events, $65.3 \pm 0.7\%$ are singles, $32.1 \pm 0.7\%$ are doubles, and triples and quadruples make up $1.0 \pm 0.1\%$ and $1.6 \pm 0.2\%$ of them, respectively. These pattern fractions are consistent with the ratios observed in the NXB slew observations, indicating that the 2--7~keV energy band of the AB~Doradus observations with the filter closed is also dominated by the unfocused background of the \xmm\ EPIC-pn. 
Case~B frames for the SNR~21.5$-$09 observations include a total of 170114 valid events and have a lower fraction of singles ($61.6\pm0.1\%$) and a larger percentage of doubles ($34.5\pm0.1\%$) is detected compared to the AB~Doradus and NXB observations.

In Case~C frames of the NXB observations, we find a total of 3622 valid events in the 2--7~keV band. The majority of the valid events ($67.8\pm0.8\%$) are singles, while doubles make up $30.1\pm~0.7 \%$ of the total events. We find that  $1.0 \pm 0.2\%$ and $1.1\pm0.2\%$ are triples and quadruples, respectively. Fractions of the valid patterns in Case~C frames are shown in Table~\ref{table:branching}. Examining AB~Doradus observations with the closed filter, we find a much lower number of valid events (a total of 322) in the Case~C frames compared to Case~B frames, consistent with the results we find in NXB observations (see Figure \ref{fig:var_frames}). Of these events $69.4 \pm 2.6\%$ are singles, $28.5\pm2.5\%$ are doubles, $1.0\pm0.6\%$ and $1.0\pm0.6\%$ of them are triples and quadruples. The branching ratios in this band are consistent with the fractions of valid event patterns observed in the slew NXB SWM data, indicating that the 2--7~keV band of AB~Doradus observations is dominated by the EPIC-pn's unfocused X-ray background, as in the Case~C frames. 

In Case~C frames in the SNR~21.9-05 observations (a total of 4663 frames), the fraction of singles is lower ($60.7\pm0.7\%$), while doubles are higher ($35.9\pm0.7\%$) compared to both NXB dominated slew and AB~Doradus observations in the source dominated hard band. Sparse statistics of triples and quadruples do not allow us to compare their branching ratios to the unfocused X-ray background.
The key result here is that the valid events that make up the instrumental background have slightly different branching ratios to celestial X-rays in the 2-7~keV band. High energy events with harder spectrum have more  probability to have high splint event ratios compared to the low energy photons with soft spectrum. Since SNR~21.9-05 has a harder spectrum, the split event ratios are expected to be higher than the split event ratios in the non X-ray background.

\subsection{Time Interval Between Valid Events}
\label{sec:branching}

\begin{figure}
\centering
\vspace{2mm}
\vspace{-2mm}\includegraphics[width=0.48\textwidth]{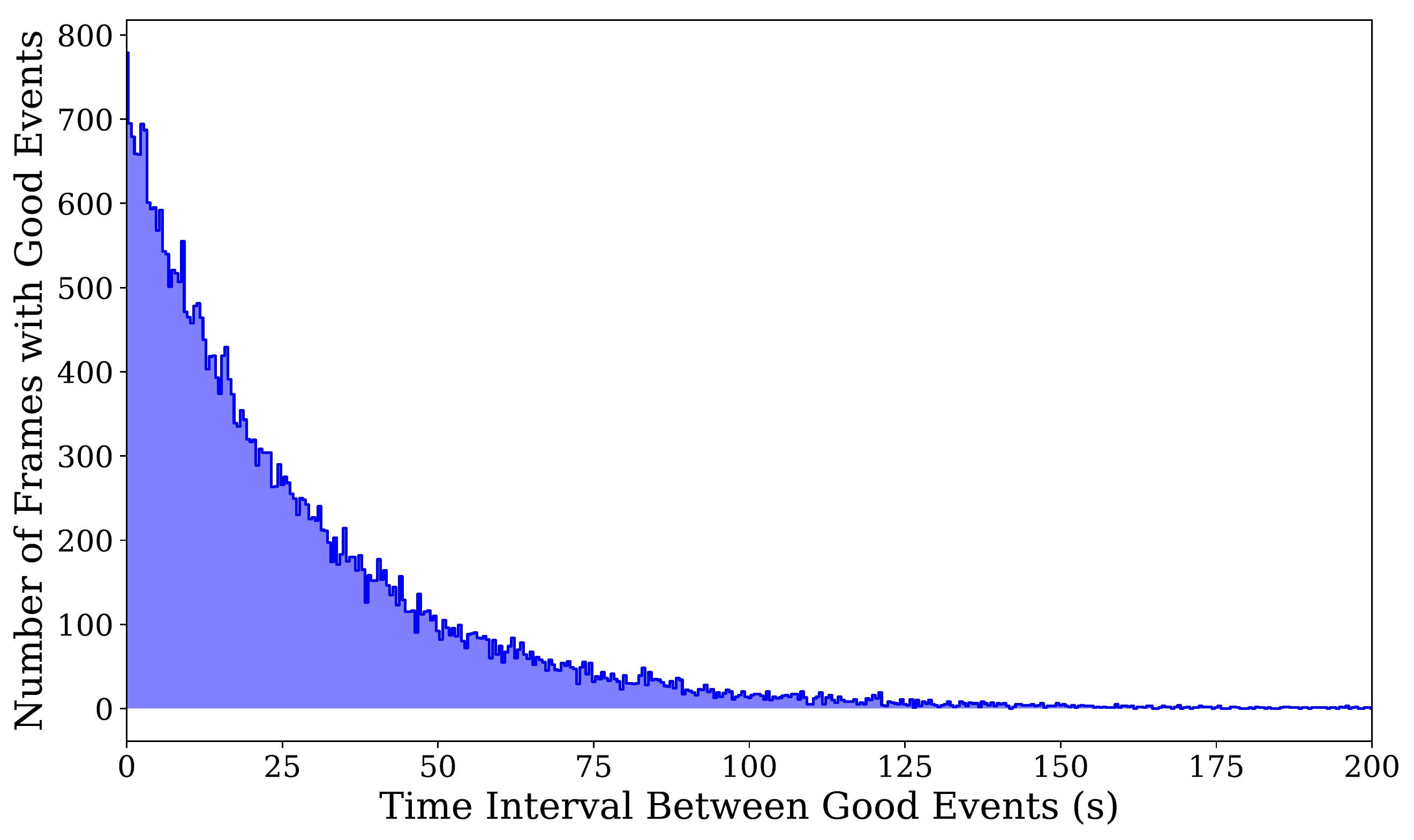}
\caption{ Distribution of the time intervals between valid events in Case~B frames showing the exponential form expected for uncorrelated events.  The exponential time constant is equal to the mean time between Case~B events.}
\label{fig:dt_valid}
\vspace{2mm}
\end{figure}

We further investigate whether there is a temporal correlation between the arrival times of valid events. The distribution of the arrival times of successive events in Case~B frames in the 2--7~keV energy band is shown in Figure \ref{fig:dt_valid}. If the valid events are independent of each other, the distribution is expected to be exponential, with a time constant close to the mean time between events (the reciprocal of the mean rate of Case~B frames). We find that the mean difference in the arrival times of the valid events in Case~B frames is 26.7~second, which is comparable to the time constant of the exponential distribution. There is no evidence of a characteristic time interval between events shorter than the mean time interval. This also confirms that the 2--7~keV band is dominated by the unfocused background and that our filtering has removed most of the instrumental artifacts associated with the long reset time constant of the output amplifiers \citep{freyberg04}. We do not find any significant departures from expectations in the arrival time of valid events depending on the particle environment.

\subsection{Spectral Properties of Valid and Particle Events}
\label{sec:spectral}
\begin{figure}
\centering
\vspace{2mm}\includegraphics[width=0.48\textwidth]{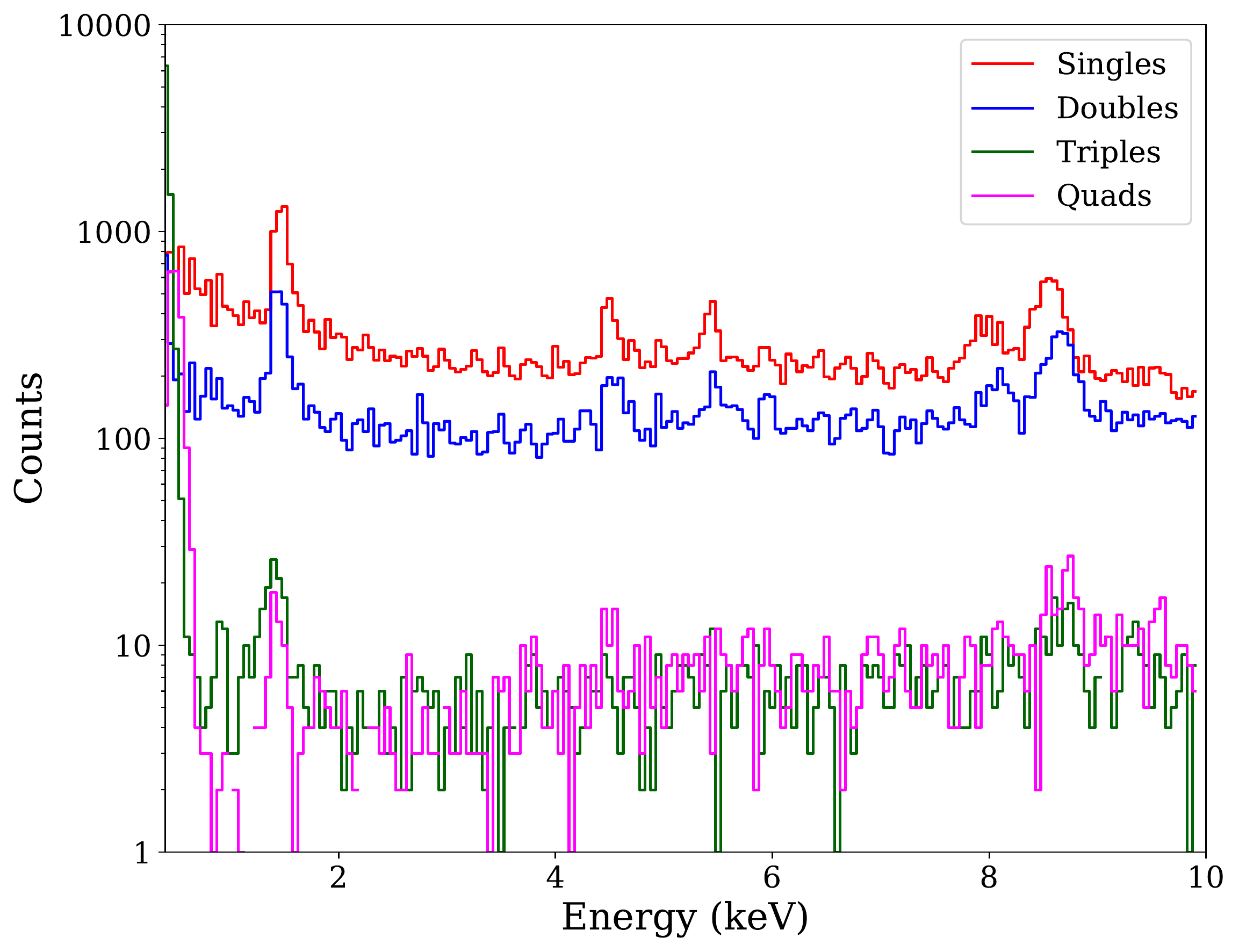}
\caption{Pulse-height spectra of valid events (single pixel events in red, doubles in blue, triples in green, and quadruples in purple) in Case~B frames observed in the \xmm\ EPIC-pn filter-wheel-closed observations. Fluorescent instrumental lines of Si K$\alpha$ (1.75~keV), Ti K$\alpha$ (4.5~keV), Cr K$\alpha$ (5.4~keV),  Cu K$\alpha$ (8.0~keV), and Zn K$\alpha$ (8.6~keV) are visible in the spectra.}
\label{fig:spec_valid}
\vspace{2mm}
\end{figure}

We first extract spectra of all valid events based on their patterns in Case~B frames (see Figure \ref{fig:spec_valid}). Overall the spectra of singles and doubles are quite flat, while the spectra of triples and doubles are slightly positively sloped towards higher energies. Additionally, we removed events that are located at the detector boundaries since it is challenging to determine if the event detected at the boundary is a single pixel event or is the partially collected charge of an event that landed off the active area of the detector.

\begin{figure*}
\centering
\includegraphics[width=0.49\textwidth]{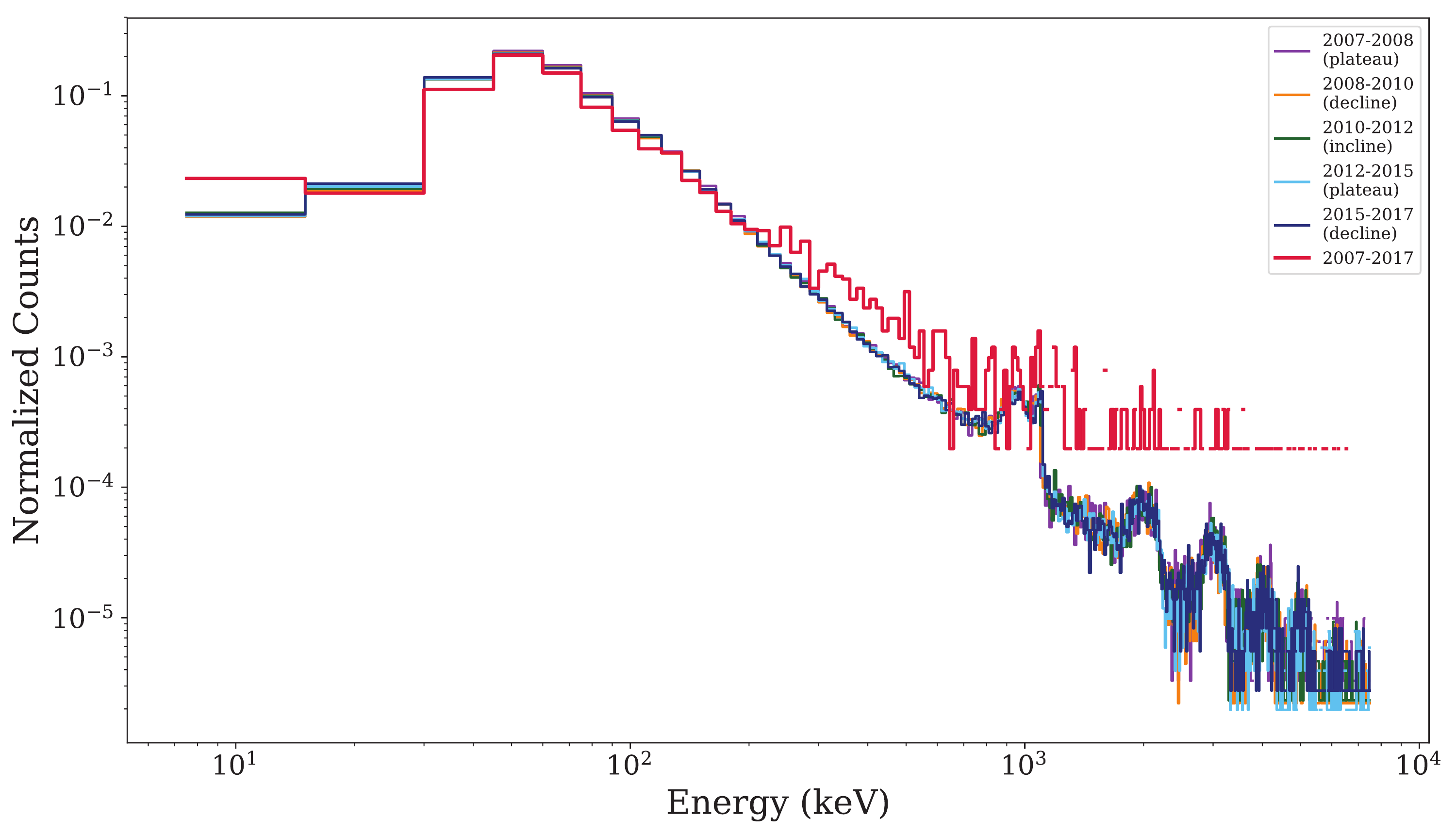}
\includegraphics[width=0.49\textwidth]{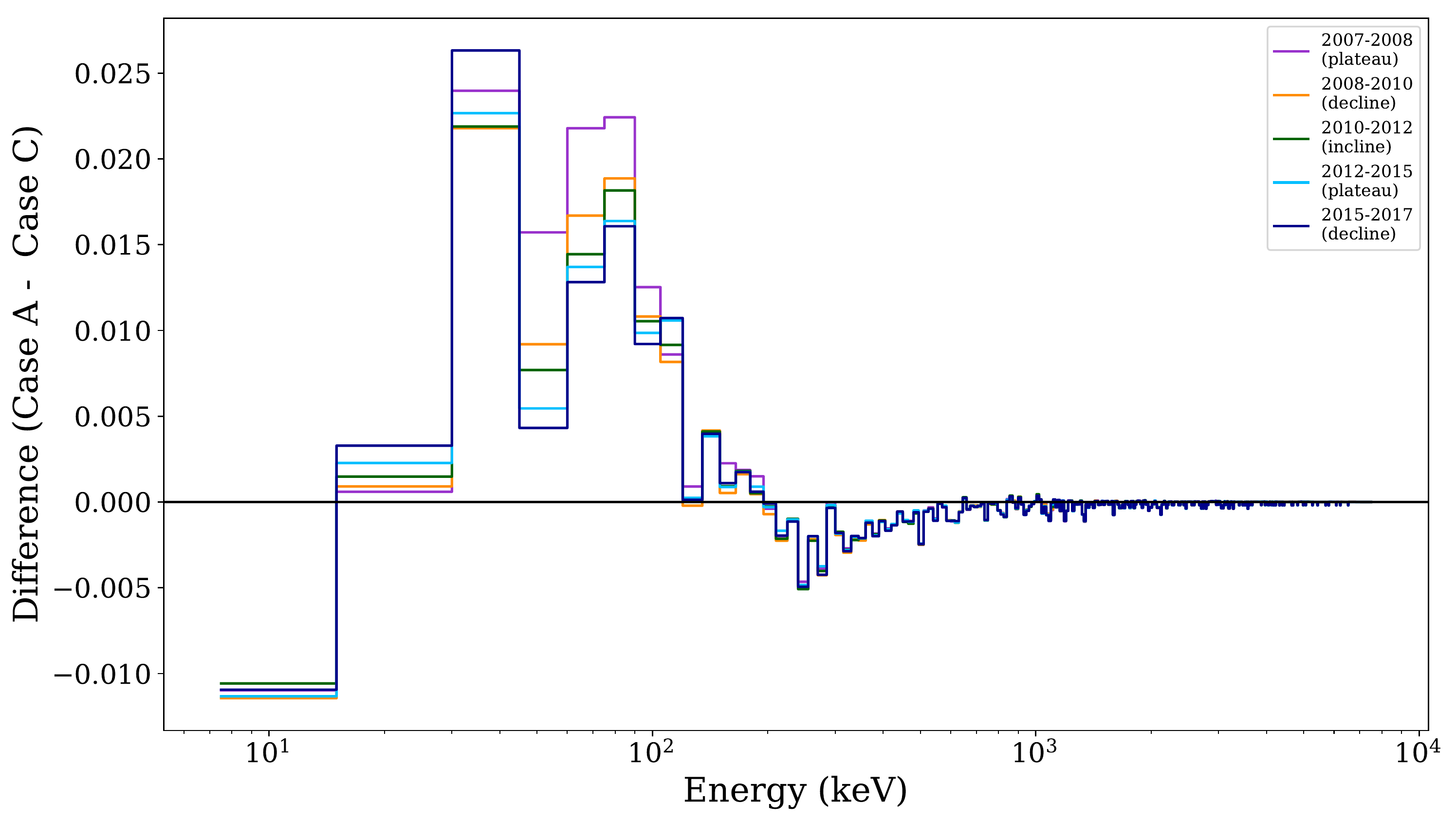}
\caption{Left Panel: Pulse-height spectra of the particle events in Case~A frames.  The data have been divided into five time intervals to sample the variation of the particle spectra with the Solar cycle and normalized by the number of frames in each period. Over-plotted in red is the pulse-height spectrum of the particle events in Case~C frames between 2007 and 2017. The periodic structure observed at high energies is an aliasing effect due to binning and ADC saturation limit. Flattening of the spectrum of the red (Case~C) histogram above $\sim$200~keV relative to the other histograms (Case~A) is clearly visible. Right Panel shows the difference between the normalized pulse height spectra of Case~A frames and Case~C frames. }
\label{fig:ene_particle}
\vspace{2mm}
\end{figure*}

Next, we investigate the spectral properties of the particle tracks in Case~A and C frames. The total energy of particle tracks is obtained by summing the charge in spatially connected pixels found by our image segmentation
algorithm. We then generate the spectra of these particle events and normalize them by the total frames in each observation (as given in Table~\ref{table:obs}). The spectra of the particle events are shown in Figure~\ref{fig:ene_particle}. The overwhelming
statistical power in Case~A frames (see Figure~\ref{fig:var_frames}) allows us to examine particle spectra in different
phases of the solar cycle: the observations in the solar
activity plateau between 2007 -- 2008 (in magenta), 2012 -- 2015 (in cyan), decline in solar activity between 2008--2010 (in orange) and 2015--2017 (dark blue), and increase in solar activity between 2010--2012 (in green). We then overplot the spectra of particle events in Case~C frames from
all epochs (2007--2017) in red in Figure~\ref{fig:ene_particle} with the same
bin size of 7~keV. Due to the limited number of Case~C
frames, we combine all particle events detected between
2007 and 2017.
Figure~\ref{fig:ene_particle} shows spectra of particle events in Case~A and Case~C frames (left) and the difference between them (right). The spectra of the particle events in Case~A frames are strikingly similar to one another and independent of the solar cycle. We find a significant difference between the spectra of particle events that are detected in Case~C frames and those detected in Case~A frames. We observe a steeper slope in the energy band $<$~200~keV in the Case~A spectra, and above $>$~200~keV the Case~C spectrum flattens. This may indicate that the particle events that create showers of valid events while passing through the detector housing (in Case~C frames) originate from a different particle population or geometry than primary particles detected in Case~A frames. The observed flattening of the spectrum of Case~C frames above 1.5~MeV is likely due to combination of limited statistics, lack of sensitivity, and ADC saturation limit.

Another implication of Figure~\ref{fig:ene_particle} is that if the spectra of particle events in the Case B and Case C frames are normalized to the counts in the 250-750~keV energy band, the spectral slopes become consistent between these two spectra. In this case, an excess of low energy particle events is observed on the Case C frames compared to the Case B frames in the lower energy band below $200$~keV. This may indicate that low energy particle events are more likely to convert into showers and create secondaries that are detected on the detector.

\begin{figure*}
\centering
\includegraphics[width=\textwidth]{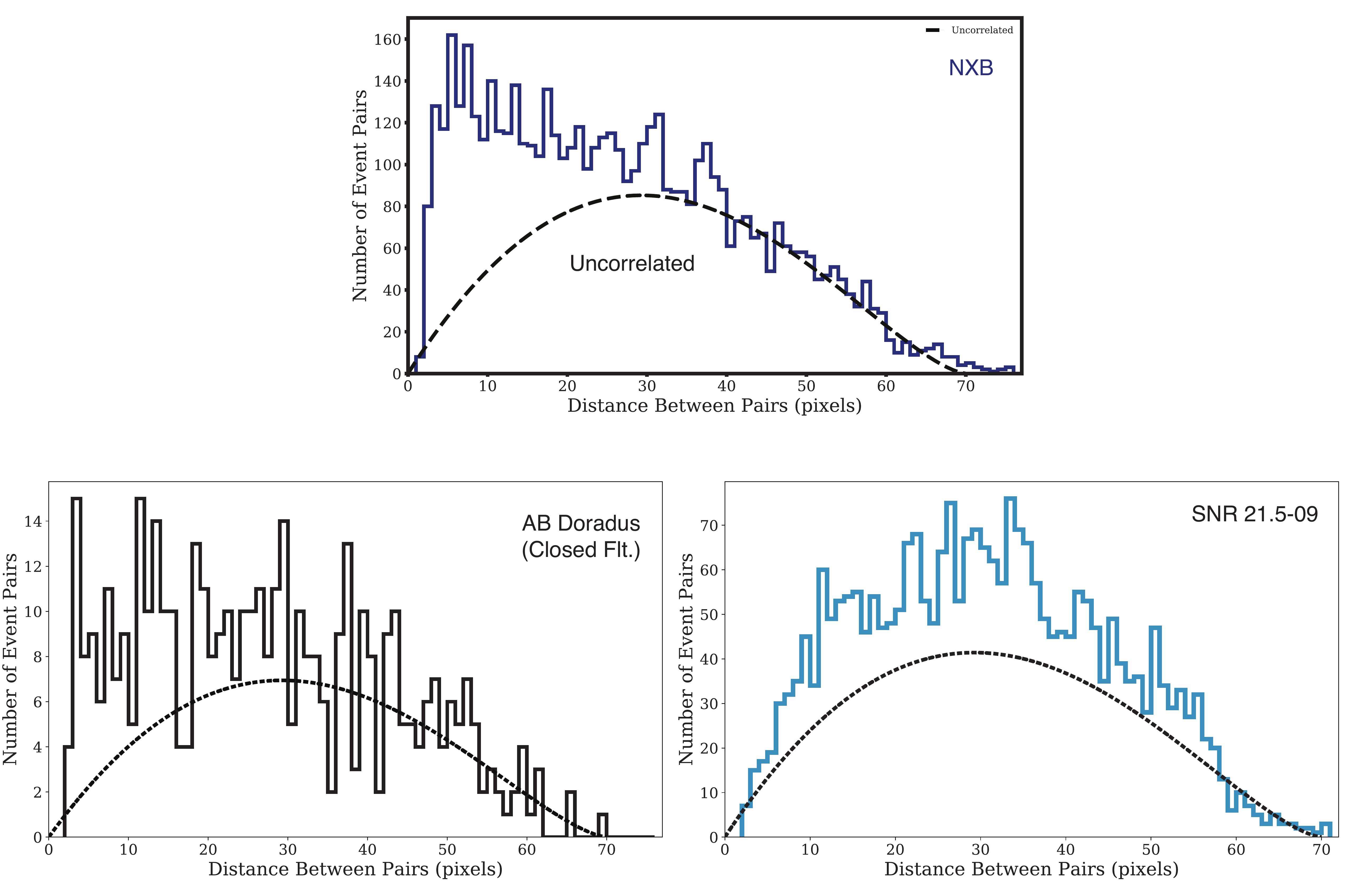}
\vspace{2mm}
\caption{The distribution of distances between valid events in the 2--7~keV energy band and the particle events detected in Case~C frames of the NXB with closed filter (in blue), SNR~21.5$-$09 with thin filter (in black), and AB~Doradus with closed filter (in cyan). The dashed curve in yellow indicates the expected distribution for uncorrelated event pairs. Valid events in the immediate $\sim30$~pixel vicinity of the particle events in NXB and AB~Doradus observations are highly correlated, indicating that the 2--7 keV band of these observations are dominated by the unfocused background of \xmm\ EPIC-pn. Although statistics of Case~C frames are limited, an evidence of spatial correlation in small spatial scales ($<$30 pixels) is visible in the AB~Doradus observations on the right panel. The two-point correlation function in the SNR~21.5$-$09 observations shows a distribution consistent with the theoretical distribution of uncorrelated events, indicating that 2--7~keV band is dominated by the emission from the supernova remnant.}
\label{fig:spatialCor}
\vspace{3mm}
\end{figure*}
\subsection{Spatial Correlation Between Particle and Valid Events}
\label{sec:spacor}

We further examine the distribution of distances between the centroid position of the valid and particle events in Case~C frames. In both cases, the centroids of the events are determined by the maximally charged pixels. In the case of the particle event islands, when highly energetic particles interact with the detector, often more than one pixel gets charged at the ADC saturation limits, i.e., 22.5~keV with MIP rejection off. In these cases, the center of the event is marked as the last saturated pixel in an event island to be read. The distribution of distance between particle events and their secondary valid events for the slew NXB observations is shown in blue in Figure~\ref{fig:spatialCor}. The form of the distribution expected for uncorrelated events in these frames is plotted as a dashed yellow curve. The significant excess of event pairs at small separations indicates that the valid events in the immediate $<$30~pixel area around the particle events are highly spatially correlated with the associated particle track. We note that, due to the small active area of the detector in the SWM observations (63 pixel $\times$64 pixel), our analysis is not sensitive to correlations at large scales. 

As a next step, we divide the valid events based on their patterns and reexamine the spatial correlations of singles, doubles, triples, and quadruples. We find a similar correlation between these patterns and particle events independent of their patterns. 
We also looked for energy dependence in the correlation between particle tracks and related events by dividing particle events into categories: particles with low-energies, $<200$~keV, and high energy particles with $>200$~keV (see Section \ref{sec:spectral}). We do not observe any differences in the spatial correlation between valid events and particle tracks as a function of energy of the particle events.

\begin{figure}
\centering
\includegraphics[width=0.48\textwidth]{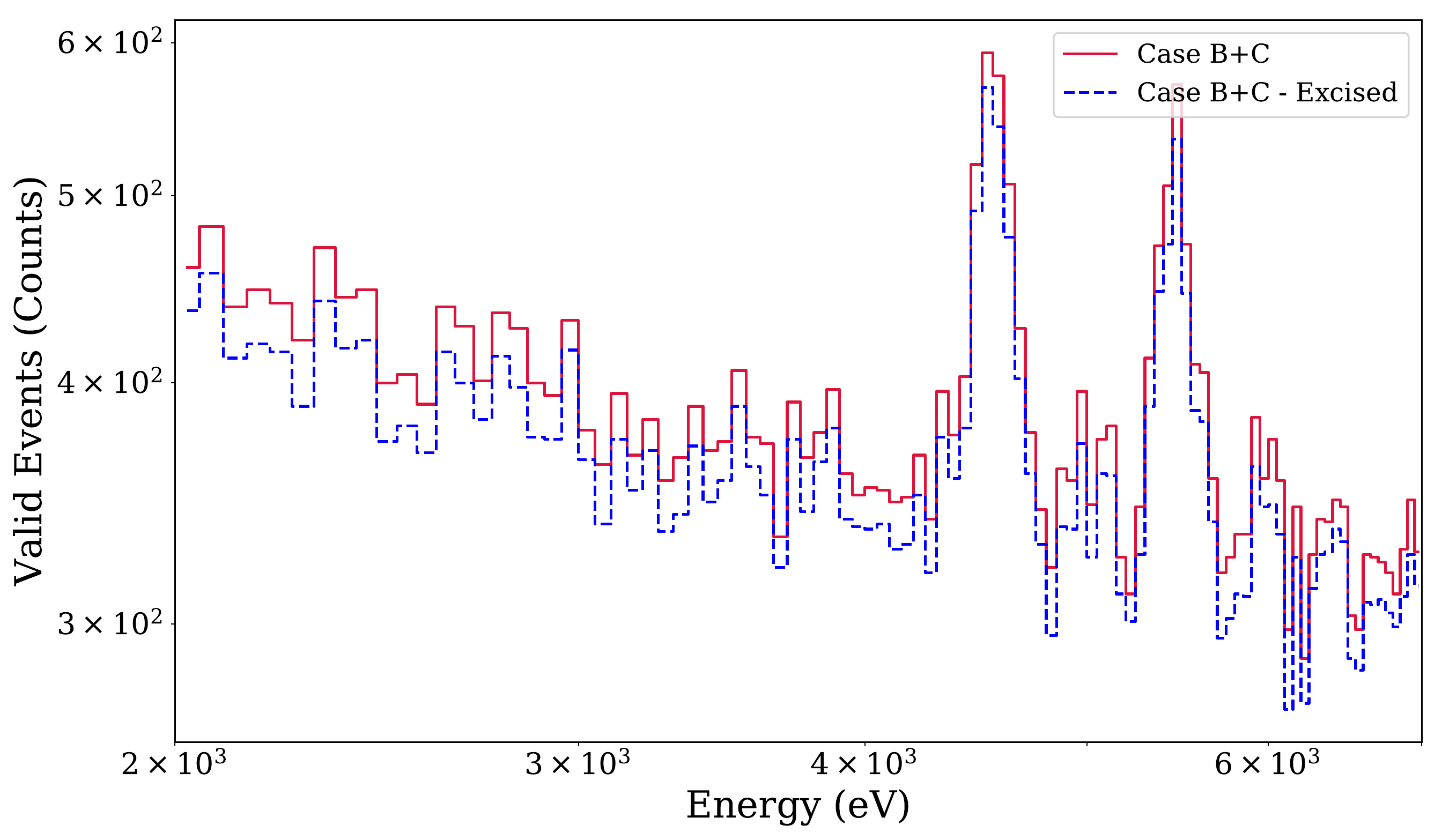}
\caption{Pulse-height spectrum of valid events detected in Case~B and C frames (in red). Rejecting valid events associated with a primary GCR by using SAC with a 30~pixels exclusion radius would reduce the particle-induced background level of the \xmm\ EPIC-pn by $\sim$10\% in the 2--7~keV energy band. The resulting background spectrum is shown in dashed blue.}
\label{fig:lc_sac}
\vspace{2mm}
\end{figure}

Next, we examine the spatial correlation between valid/particle event pairs observed in Case~C frames for the pointed observations through the closed filter of the AB~Doradus star system (black distribution in Figure~\ref{fig:spatialCor}). As expected, this histogram is similar to the one for the slew NXB observations (e.g., there is a significant excess at small spatial scales up to 30~pixels) indicating that the 2--7~keV band is dominated by the unfocused X-ray background. 

The shape of the spatial correlation between valid events and particle tracks for the SNR~21.5-0.9 data (cyan in Figure \ref{fig:spatialCor}) is similar to the form expected for uncorrelated events (yellow curve). However, the distribution of separations is more peaked than expected for pairs of randomly distributed events. This is because, although a particle is equally likely to land anywhere on the detector, the supernova is centered on the chip, causing the distribution of source photons to be peaked there. This indicates that the 2--7~keV energy band for the SNR~21.5$-$09 observations is dominated by photons from the supernova remnant and the unfocused X-ray background is subdominant. 

These results are the basis of the self-anti-coincidence (SAC) method, used to reduce secondary events associated with a particle primary. This method of partial vetoing of valid events around particle tracks shows promise at reducing the systematic error produced by the instrumental background at the expense of eliminating events from real source X-rays (based on private communication with Silvano Molendi). We find that by eliminating events that fall within 30 pixels of the peak of a particle track, the particle-induced background of the \xmm\ EPIC-pn can be reduced by $\sim$10\% (see Figure \ref{fig:lc_sac}). The results summarized here from the \xmm\ EPIC-pn FWC observations can be used to reduce the particle background level of the future silicon-based X-ray detectors. For instance, the earlier EPIC MOS results were used to optimize FWC rotation strategy to sample particle background component during \athena\ WFI observations of faint objects \citep{gastaldello17, vonkienlin18}.

\begin{figure*}
\centering
\includegraphics[width=1.\textwidth]{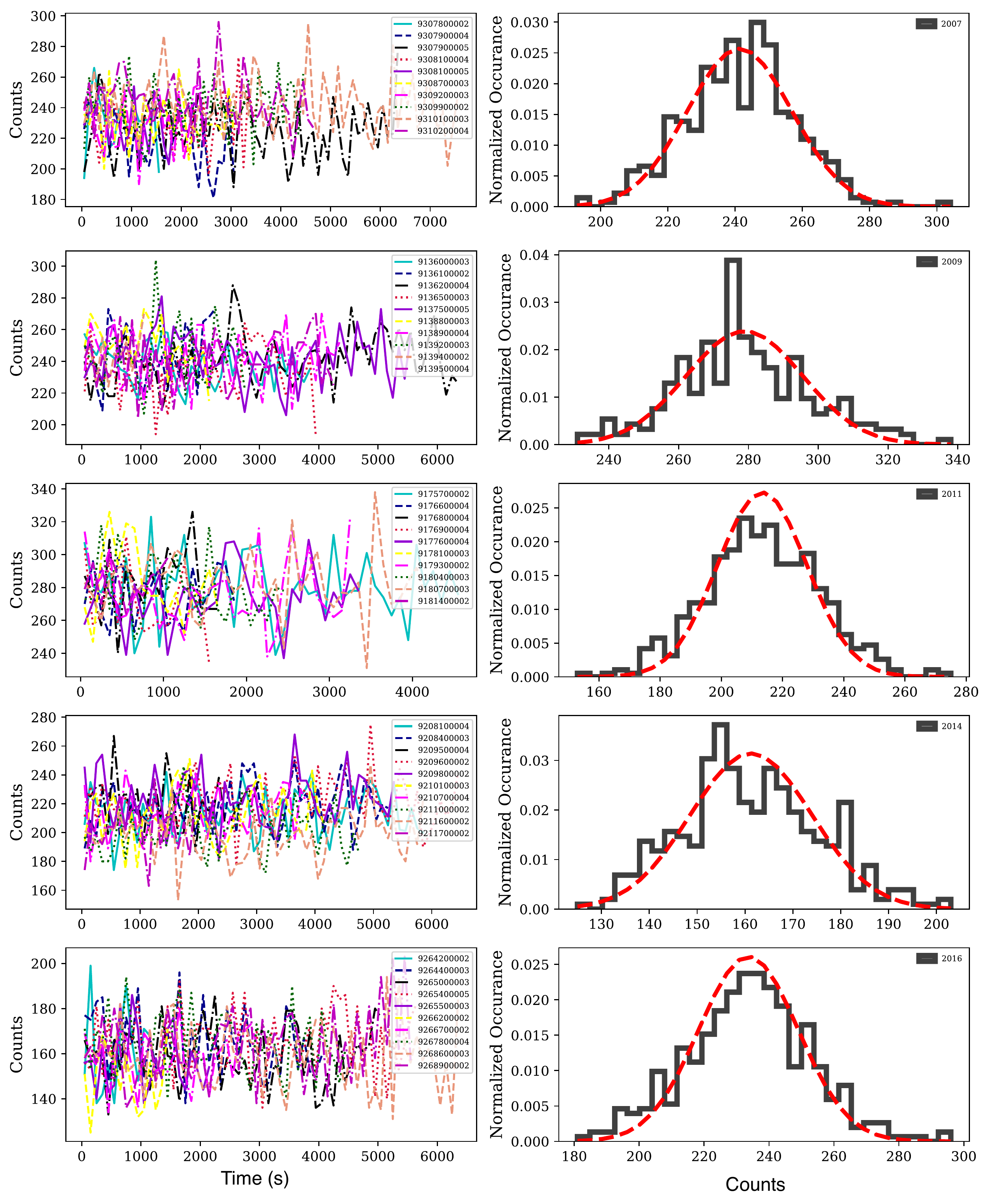}
\caption{The light curves of particle events detected in Case~A frames per 100~s binning for five different epochs in the solar cycle are shown on the left panels. The dashed lines in the right hand panels show the expected Poisson distributions around the mean. The distributions of count rates of particle events in these observations in each epoch are shown on the right panels. }
\label{fig:lc_part}
\vspace{2mm}
\end{figure*}
\begin{figure*}
\centering
\includegraphics[width=1.\textwidth]{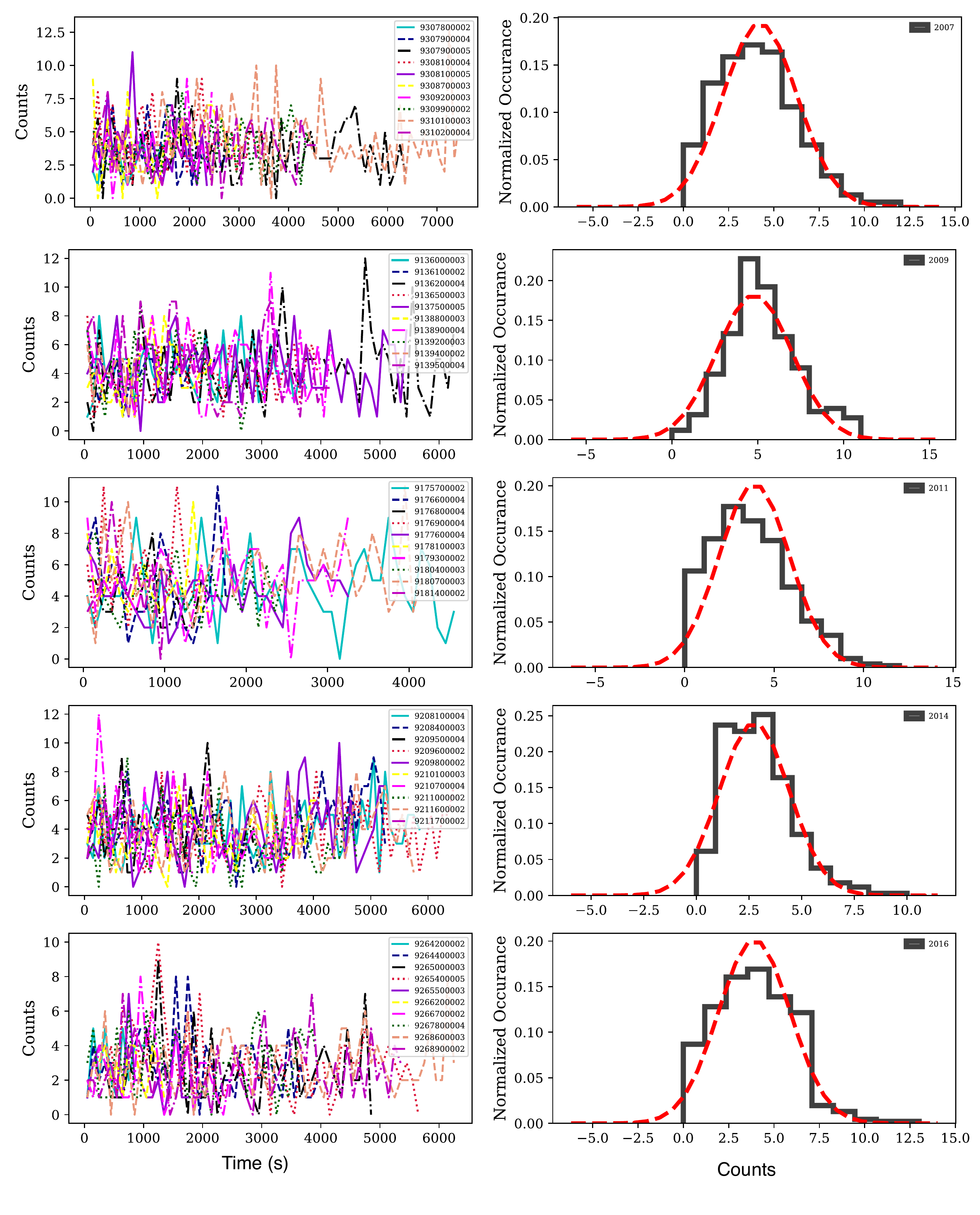}
\caption{The light curves of valid events in 2 to 7 keV band detected in Case~B frames per 200s binning for five different epochs in the solar cycle are shown on the left panels. The dashed lines in the right hand panels show the expected Poisson distributions around the mean. The distributions of count rates of particle events in these observations in each epoch are shown on the right panels. }
\label{fig:lc_valid}
\vspace{2mm}
\end{figure*}
\begin{figure}
\centering
\hspace{-2mm}\includegraphics[width=0.49\textwidth]{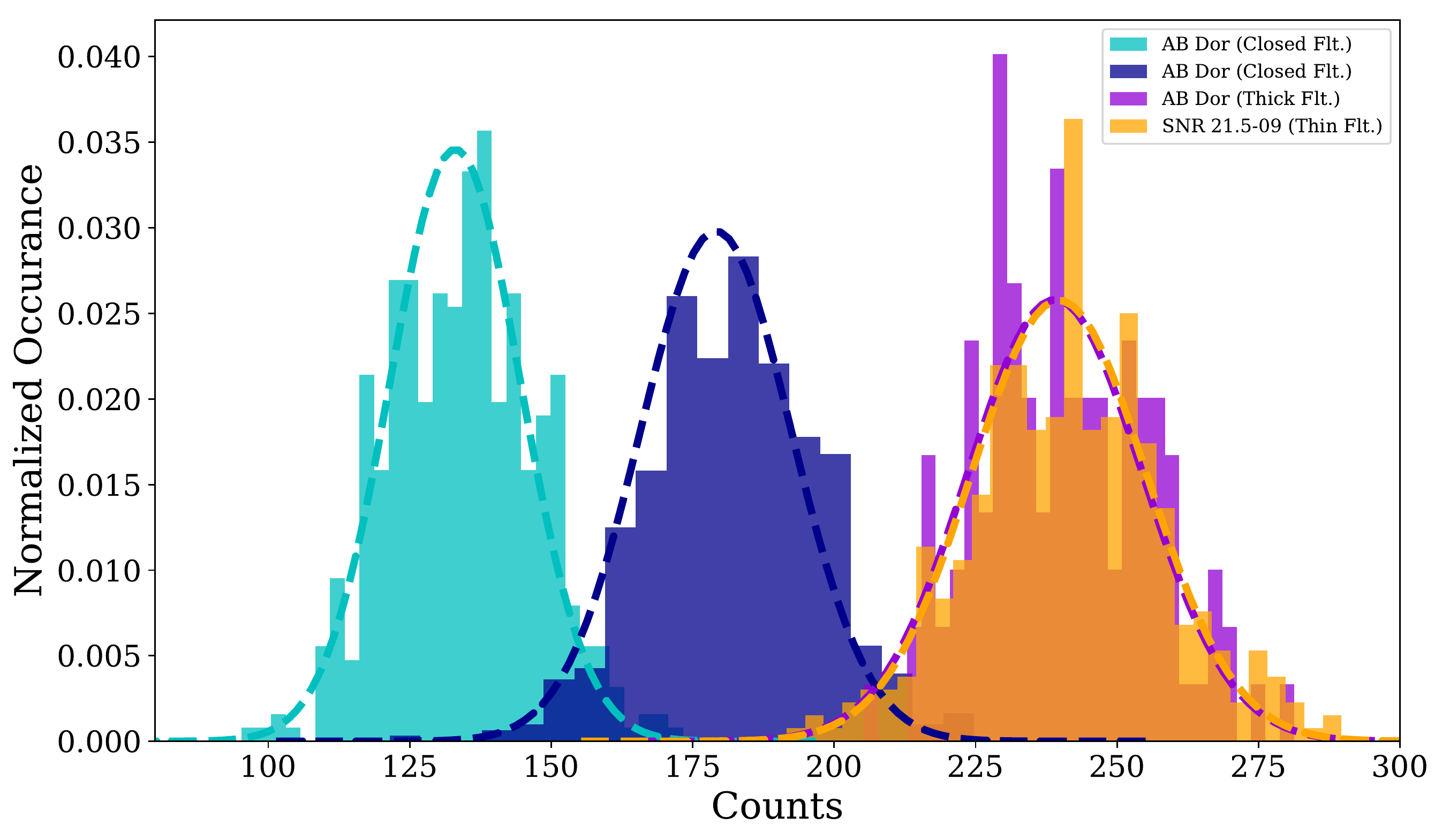}
\caption{Distribution of particle event counts per 100~s bin obtained from light curves of the pointed observations of the AB~Doradus star system taken in two different filter configurations; filter closed and with the thick filter. The observations were taken in 2002, 2004, and 2017. Comparing mean count rates with the count rates observed in NXB data (see Table \ref{table:mean_part}), we can infer that the filter closed observations were taken during solar maximum, while the 2017 observations were performed during solar minimum. The unfocused background level measured in the FWC data and solar activity are closely correlated.}
\label{fig:lc_part_pointed}
\vspace{2mm}
\end{figure}
\subsection{Short Term Variability of the Particle Environment}
\label{sec:varib}

Owing to the long term coverage of the \xmm\ EPIC-pn slew observations with the filter-wheel closed between years 2007 and 2017, we are able to probe short term variability of particle events in Case~A frames. We examine the light curves of particle events in Case~A frames in five epochs defined by the phase of the solar cycle as shown in Figure~\ref{fig:var_frames} (plateau in 2007, solar minimum in 2009, increase in solar activity in 2011, solar maximum in 2014, decrease in solar activity in 2016). The variability of the rate of particle events in Case~A frames in 10 observations taken close together in time, with 100~s binning, is shown in Figure \ref{fig:lc_part} for each epoch. We note that the observations used in producing light curves in this section differ from the observations used to generate the spectra in Section \ref{sec:spectral}. The mean, standard deviation, and skewness of the light curve counts of these particle events are given in Table \ref{table:mean_part}. The dashed curves indicate a normal distribution with Poisson standard deviation. There is no statistically significant variability of the particle tracks in any epoch.

\begin{table*}
\centering
\caption{Statistics of Light Curves of Particle Event Rates in Case~A Frames binned for 100s}
\label{table:mean_part}
\begin{tabular}{lcccccccc}
\hline\hline

Date & Epoch & Mean & Std Dev. & Skewnes & KS test   \\
    &   & & & & D-value & p-value\\
\hline
\hline
2007 & Plateau & 238 & 16 & 0.01 & 0.16  & 0.76\\
2009 & Solar Miminum & 276 & 19 & 0.23 & 0.13 & 0.93 \\
2011 & Solar Activity Increase & 210 & 18 & 0.06 & 0.10 & 0.99\\
2014 & Solar Maximum &  159 & 14 & 0.24 & 0.13  & 0.93\\
2016 & Solar Activity Decrease & 230 & 18 & -0.01 & 0.13  & 0.93\\
2002 & AB~Doradus (Closed Flt.) & 133 & 12 & 0.08 & 0.17 & 0.76\\ 
2004 & AB~Doradus (Closed Flt.) & 179 & 15 & -0.82 & 0.25 & 0.78 \\
2017 & AB~Doradus (Thick Flt.) & 239 & 15 & 0.25 & 0.13 & 0.94\\ 
2017 & SNR 21.5$-$09 (Thin Flt.) & 239  & 16 & 0.05 &0.16 & 0.76 \\

\hline\hline
\end{tabular}
\end{table*}

We find that the mean values of the particle counts vary between 133 and 276, depending on the solar activity (see Table \ref{table:mean_part}); however, the standard deviations (12-19) are remarkably small and independent of solar cycle. The mean of particle event counts observed per 100~s can be as high as 276 during solar minimum, while it can be as low as 133 during solar maximum. In general, each distribution is well-matched to the Poisson distribution expected for a constant mean rate (shown in dashed curves). We do not observe significant irregularities or outlier particle events in the light curves. 

Similarly, we examine the light curves of particle events in Case~A frames of the pointed \xmm\ EPIC-pn SWM observations of AB~Doradus and SNR~21.5$-$09. The histograms of the light curves are similarly tightly distributed around the mean as observed in NXB observations, close to the expected Poisson distribution. The AB~Doradus observations were taken in 2002 and 2004, and the observed mean values are 133 and 179 while the Sun was active. The AB~Doradus observations with thick filter were taken in 2017, when the solar activity was approaching its minimum, therefore a mean rate of 239 particle tracks is observed during those observations. These count rates are consistent with the count rates we observe in NXB dominated slew observations. The SNR~21.5$-$09 observations were also taken in 2017 while solar activity was approaching minimum. The observed mean value of 239 indicates that these observations were performed when the solar activity was at minimum. 

To further test the similarities in the background light curves against the Poisson distribution, we computed Kolmogorov-Smirnov statistics. The Kolmogorov-Smirnov test determines the probability of two samples being drawn from the same distribution. The high values of probabilities ($>0.73$) indicate that these background light curves distributions are originating from the same underlying Poisson distribution.

We also show distributions of the number of valid events in the 2--7~keV energy band in Case~B frames in Figure~\ref{fig:lc_valid}, binned for longer time intervals (200~s) to allow for the lower event rate. Similarly, we do not observe any significant deviations from Poisson distributions for the numbers of particle related events in the filter-wheel closed data.

\section{Conclusions}
\label{sec:concl}
In this work, we present analysis of the unfocused X-ray background of the \xmm\ EPIC-pn operating in small window mode with a fast frame time. These observations were taken while \xmm\ was slewing to a variable source that was to be observed in SWM, while the filter-wheel was in the closed position and the MIP rejection algorithm was turned off. This dataset uniquely allows us to study temporal, spectral, and spatial properties of particle primaries and their secondaries generated as a result of the interactions with the detector housing, which constitute the unfocused instrumental background for the science observer. We also compare our results from the unfocused background, NXB, with the pointed filter closed observations of a star system AB~Doradus and observations of a supernova remnant SN~21.5$-$09 taken with the thin filter. Owing to the large number of frames, we were able to independently study the frames with just primary particles (Case~A), with just secondary valid events (Case~B), and with both primary particle and secondary valid events (Case~C). Our major results are as follows.

\begin{itemize}
    \item Examining the branching ratios of event morphologies, we find that the vast majority of valid events in Case~B frames of NXB observations are single pixel events (65.6~$\pm$~0.2\%) and double pixel events (31.3~$\pm$~0.2\%). Comparing these ratios with the observations of a supernova remnant, we find that in both cases, Case~B frames have a significantly smaller fraction of singles (61.6~$\pm$~0.1), and larger fraction of doubles (34.5~$\pm$~0.1\%). The fraction of singles in Case~C frames of the unfocused NXB (67.8~$\pm$~0.8\%) is also higher compared to that in the supernova observations (60.7~$\pm$~0.7\%). In both cases, the differences are statistically significant. That is, the valid events in the instrumental background have somewhat different branching ratios than those of the celestial X-rays.

    \item The mean difference between the observed arrival times of successive valid events in Case~B frames matches the reciprocal of the event rate, as expected.  We do not observe any structure in the distribution of the time intervals suggestive of a temporal correlation between background events, or detector or background effects on the time interval between valid events. As expected, all background events appear to be independent and uncorrelated.

    \item The energy spectrum of the particle tracks in frames with valid events is somewhat flatter than that of the tracks in frames with no valid events. This result indicates that the particle events detected with secondary events in the same frame (Case~C) might be due to a different population of particles passing through, or a different geometry compared to the primary particle events that do not generate secondary showers in the detector housing. We also found that when the spectra of particle events are normalised to the high energy band (250--750~keV), an excess of low energy particle are observed in the Case~C frames (frames with at least one primary and secondary particles) compared to the Case~B frames (frames with just primaries). This may indicate that low energy particles are more likely to interact with the detector housing and create secondary particle showers. 
    
    \item We find a significant spatial correlation between particle and valid events in Case~C frames on small spatial scales up to 30~pixels (4500~microns) of the unfocused background observations with the closed filter in the 2--7~keV band. In the observations of the supernova remnant SNR~21.5$-$09 no spatial correlation between the valid events and particle events is observed, as expected. Rejecting valid events (``self-anticoincidence''' or ``SAC'')  within 30~pixels around the primary GCRs reduces the absolute level of the particle-induced background of \xmm\ EPIC-pn by $\sim$10\% in the 2--7~keV energy band.
    
    \item Light curves of particle events in Case~A frames display a tight distribution, with mean particle counts of 133 to 276 per 100~s bin, depending on the phase of the solar cycle. The mean number of particle events per 100~s bin can be as high as 276 during solar minimum, while it can be as low as 133 during solar maximum. The sample standard deviations of the count per 100~s bin are consistent with expectations for Poisson distributions with the observed means. There is no evidence for any short-term temporal variability in the GCR component of the instrumental background, beyond what is expected for Poisson noise. KS-test results indeed indicate that the distribution of count rates in the light curves of Case A and Cade B frames are consistent with the Poissonian distribution around the mean rate. These means and distributions can be used to monitor particle rates and estimate the level of unfocused background of future X-ray imaging detectors.
    
    \item Light curves of valid events (secondaries generated by primary GCRs) also display a tight distribution around the mean, consistent with the expected Poissonian distribution. Similarly, there is no significant evidence for any short-term temporal variability in the secondary background events. These observed rates closely correlate with the solar cycle and particle rates and can be used to predict the level of unfocused X-ray background.

\end{itemize}

Similar analyses of the unfocused component of the X-ray detector background have been performed on the {\it Chandra} stowed ACIS data and the Neil Gehrels Swift Telescope XRT data \citep[e.g.,][]{bartalucci14, grant18, bulbul18}. Results we present in this work should help to understand and reduce the particle background level in other Si-based X-ray detectors (e.g., the Wide Field Imager on \athena\ and the \erosita\ instrument on board the Spectrum Roentgen Gamma observatory). The SWM frame time of 5.67~ms is similar to the \athena\ WFI default frame, allowing us to validate GEANT4 simulations of the \athena\ WFI unfocused background \citep[see][]{miller20}.

Beyond validating the GEANT4 simulations for the \athena\ WFI, this study also lays the ground work for application of self-anticoincidence to reduce the unfocused background in silicon-based X-ray detectors, e.g. WFI on board of \athena\  \citep{nandra13}, \erosita\ on board of SRG \citep{merloni2012}, EPIC on board of \xmm\ \citep{janse01}, and HDXI on board of {\it Lynx}  \citep{gaskin19}. The results obtained from this work will be used to develop both on-board and ground-based algorithms to better characterize and improve background rejection for silicon-based X-ray imaging detectors. The self-anticoincidence method, and the results presented in this work, will help reduce the Athena WFI particle background and increase the signal-to-noise in background-dominated observations, such as galaxy cluster outskirts and deep surveys, enhancing the science return of \athena.

\section*{Acknowledgements}
%
Authors thank the anonymous referee for helpful comments on the draft. We gratefully acknowledge support from NASA grant NNX17AB07G, administered by Penn State, and from NASA contracts NAS 8-37716 and NAS 8-38252. 

This paper made use of the simulations from \texttt{GEANT} software \citep{tenzer2010} and \xmm\ SAS analysis software \citep{gabriel2004}. This work made use of SciPy \citep{jones_scipy_2001}, matplotlib, a Python library for publication quality graphics \citep{Hunter:2007}, Astropy, a community-developed core Python package for Astronomy \citep{2013A&A...558A..33A}, NumPy \citep{van2011numpy}. 

\appendix
\section{Observations}
\startlongtable
\begin{deluxetable*}{ccccccccccc}\tabletypesize{\scriptsize}
\tablecaption{\xmm\ EPIC-pn Small Window Mode Observations Taken in the Filter-Wheel Closed Set-up during slewing phase. \label{table:obs}}
\tablehead{
\colhead{Obs. } & \colhead{Obs. ID} & \colhead{Revolution}&\colhead{Exposure}& \colhead{Number of } &\colhead{} &\colhead{Obs.} & \colhead{Obs. ID} & \colhead{Revolution}&\colhead{Exposure}& \colhead{Number of }\\
\colhead{Index} & \colhead{} & & \colhead{(ks)} & Frames &\colhead{}  &\colhead{Index} & \colhead{} & & \colhead{(ks)} & Frames\\
}
\startdata
0	&	9136000003	&	1360	&	3.91	&	690114	&	&	155	&	9213400002	&	2134	&	3.4	&	599994	\\
1	&	9136100002	&	1361	&	2.37	&	418658	&	&	156	&	9214900004	&	2149	&	1.42	&	249893	\\
2	&	9136200004	&	1362	&	6.42	&	1132080	&	&	157	&	9217500002	&	2175	&	1.85	&	326346	\\
3	&	9136500003	&	1365	&	4.07	&	717519	&	&	158	&	9218200004	&	2182	&	2.58	&	454189	\\
4	&	9137500005	&	1375	&	5.65	&	995629	&	&	159	&	9218300002	&	2183	&	2.47	&	435767	\\
5	&	9138800003	&	1388	&	2.24	&	395752	&	&	160	&	9219200004	&	2192	&	3.13	&	551326	\\
6	&	9138900004	&	1389	&	4.42	&	778932	&	&	161	&	9223300002	&	2233	&	5.31	&	935852	\\
7	&	9139200003	&	1392	&	2.86	&	504355	&	&	162	&	9223700003	&	2237	&	2.7	&	475590	\\
8	&	9139400002	&	1394	&	1.15	&	201914	&	&	163	&	9225900003	&	2259	&	3.18	&	560297	\\
9	&	9139500004	&	1395	&	4.31	&	759510	&	&	164	&	9226100002	&	2261	&	5.88	&	1035931	\\
10	&	9139700002	&	1397	&	2.18	&	383899	&	&	165	&	9226400004	&	2264	&	2.63	&	464485	\\
11	&	9140100004	&	1401	&	1.37	&	241883	&	&	166	&	9227500006	&	2275	&	3.48	&	614237	\\
12	&	9141000003	&	1410	&	4.08	&	719023	&	&	167	&	9227600002	&	2276	&	2.24	&	394874	\\
13	&	9142800004	&	1428	&	5.71	&	1005950	&	&	168	&	9229000002	&	2290	&	1.84	&	324827	\\
14	&	9143300002	&	1433	&	3.71	&	654472	&	&	169	&	9229600003	&	2296	&	1.84	&	323646	\\
15	&	9144300004	&	1443	&	6.17	&	1087659	&	&	170	&	9229700003	&	2297	&	1.52	&	267659	\\
16	&	9144500003	&	1445	&	2.97	&	522981	&	&	171	&	9229900003	&	2299	&	1.51	&	265961	\\
17	&	9144700003	&	1447	&	0.99	&	175325	&	&	172	&	9230900003	&	2309	&	4.29	&	756196	\\
18	&	9144900005	&	1449	&	1.33	&	234024	&	&	173	&	9231800002	&	2318	&	3.82	&	673051	\\
19	&	9145700003	&	1457	&	5.2	&	917470	&	&	174	&	9232100004	&	2321	&	6.81	&	1199982	\\
20	&	9146300006	&	1463	&	3.33	&	587072	&	&	175	&	9233000003	&	2330	&	2.29	&	403991	\\
21	&	9147500002	&	1475	&	1.88	&	331127	&	&	176	&	9233200003	&	2332	&	3.94	&	693939	\\
22	&	9147900002	&	1479	&	1.86	&	327992	&	&	177	&	9233900005	&	2339	&	4.44	&	782505	\\
23	&	9148000004	&	1480	&	1.28	&	225991	&	&	178	&	9236300002	&	2363	&	3.8	&	669151	\\
24	&	9148400003	&	1484	&	2.5	&	440018	&	&	179	&	9236600002	&	2366	&	5.91	&	1042100	\\
25	&	9149500002	&	1495	&	4.75	&	838019	&	&	180	&	9236700002	&	2367	&	6.49	&	1144493	\\
26	&	9151000002	&	1510	&	6.67	&	1175239	&	&	181	&	9236900002	&	2369	&	4.91	&	865822	\\
27	&	9151000003	&	1510	&	3.59	&	632906	&	&	182	&	9238200002	&	2382	&	5.89	&	1038778	\\
28	&	9151300002	&	1513	&	5.52	&	973663	&	&	183	&	9238700004	&	2387	&	4.35	&	766698	\\
29	&	9151600004	&	1516	&	1.98	&	349686	&	&	184	&	9239400003	&	2394	&	3.85	&	678823	\\
30	&	9151700002	&	1517	&	0.95	&	168322	&	&	185	&	9240900002	&	2409	&	3.38	&	595051	\\
31	&	9152300002	&	1523	&	2.14	&	377646	&	&	186	&	9241200002	&	2412	&	2.11	&	371796	\\
32	&	9152400002	&	1524	&	3.32	&	585622	&	&	187	&	9241500002	&	2415	&	5.36	&	944903	\\
33	&	9152700003	&	1527	&	3.79	&	667745	&	&	188	&	9241600002	&	2416	&	1.11	&	196163	\\
34	&	9152900002	&	1529	&	3.54	&	624743	&	&	189	&	9242200003	&	2422	&	2.04	&	359330	\\
35	&	9153000003	&	1530	&	4.17	&	735806	&	&	190	&	9242700002	&	2427	&	2.28	&	401729	\\
36	&	9153100004	&	1531	&	5.72	&	1007747	&	&	191	&	9243000003	&	2430	&	3.56	&	628516	\\
37	&	9153200003	&	1532	&	5.31	&	936525	&	&	192	&	9245700004	&	2457	&	2.3	&	405332	\\
38	&	9153300002	&	1533	&	4.7	&	827869	&	&	193	&	9247900002	&	2479	&	1.03	&	182172	\\
39	&	9153400002	&	1534	&	3.7	&	652071	&	&	194	&	9248700002	&	2487	&	1.23	&	216006	\\
40	&	9153400004	&	1534	&	4.92	&	868326	&	&	195	&	9248900002	&	2489	&	2.41	&	425183	\\
41	&	9153600002	&	1536	&	5.47	&	964802	&	&	196	&	9248900003	&	2489	&	3.63	&	640579	\\
42	&	9153600003	&	1536	&	6.18	&	1090252	&	&	197	&	9249100002	&	2491	&	6.75	&	1189397	\\
43	&	9153900002	&	1539	&	1.05	&	184286	&	&	198	&	9249300002	&	2493	&	6.28	&	1106857	\\
44	&	9154200004	&	1542	&	6.26	&	1103389	&	&	199	&	9249400002	&	2494	&	3.42	&	603517	\\
45	&	9154300003	&	1543	&	2.9	&	510531	&	&	200	&	9249500003	&	2495	&	6.09	&	1073205	\\
46	&	9154400005	&	1544	&	3.06	&	538903	&	&	201	&	9249600002	&	2496	&	3.48	&	614271	\\
47	&	9154600005	&	1546	&	4.39	&	773231	&	&	202	&	9249700002	&	2497	&	3.49	&	615429	\\
48	&	9156800003	&	1568	&	5.83	&	1027723	&	&	203	&	9249800002	&	2498	&	1.72	&	302452	\\
49	&	9158100002	&	1581	&	4.66	&	822194	&	&	204	&	9249900002	&	2499	&	4.02	&	708999	\\
50	&	9158900004	&	1589	&	3.34	&	589018	&	&	205	&	9252900006	&	2529	&	0.98	&	171979	\\
51	&	9160000002	&	1600	&	3.28	&	578523	&	&	206	&	9253300003	&	2533	&	3.06	&	540308	\\
52	&	9160700004	&	1607	&	1.44	&	253915	&	&	207	&	9254400002	&	2544	&	1.25	&	220908	\\
53	&	9160800004	&	1608	&	5.82	&	1025948	&	&	208	&	9254500004	&	2545	&	4.1	&	723228	\\
54	&	9160900002	&	1609	&	2.95	&	519505	&	&	209	&	9254600004	&	2546	&	4.71	&	830219	\\
55	&	9161000002	&	1610	&	0.95	&	167732	&	&	210	&	9256500002	&	2565	&	5.84	&	1028961	\\
56	&	9161300002	&	1613	&	1.08	&	189574	&	&	211	&	9256600002	&	2566	&	4.34	&	764502	\\
57	&	9161500004	&	1615	&	2.03	&	358285	&	&	212	&	9257300002	&	2573	&	1.56	&	274811	\\
58	&	9161600002	&	1616	&	6.03	&	1062547	&	&	213	&	9258700002	&	2587	&	4.26	&	751265	\\
59	&	9161900002	&	1619	&	1.53	&	268946	&	&	214	&	9258800002	&	2588	&	3.16	&	557816	\\
60	&	9162100003	&	1621	&	3.5	&	617167	&	&	215	&	9259300002	&	2593	&	5.78	&	1018257	\\
61	&	9163100002	&	1631	&	1.93	&	341051	&	&	216	&	9261200003	&	2612	&	5.39	&	950948	\\
62	&	9164900002	&	1649	&	2.56	&	450613	&	&	217	&	9261800002	&	2618	&	5.7	&	1004257	\\
63	&	9164900003	&	1649	&	3.04	&	536480	&	&	218	&	9262500003	&	2625	&	5.94	&	1046693	\\
64	&	9165500004	&	1655	&	2.32	&	408867	&	&	219	&	9263300002	&	2633	&	1.54	&	271100	\\
65	&	9166200003	&	1662	&	0.99	&	175344	&	&	220	&	9264200002	&	2642	&	1.24	&	219011	\\
66	&	9168100003	&	1681	&	1.48	&	260533	&	&	221	&	9264400003	&	2644	&	4.18	&	737026	\\
67	&	9169500002	&	1695	&	1.39	&	245523	&	&	222	&	9265000003	&	2650	&	5.15	&	908458	\\
68	&	9169600003	&	1696	&	1.55	&	273414	&	&	223	&	9265400005	&	2654	&	5.74	&	1012829	\\
69	&	9169700004	&	1697	&	3.36	&	591780	&	&	224	&	9265500003	&	2655	&	2.25	&	396033	\\
70	&	9169800002	&	1698	&	4.11	&	725288	&	&	225	&	9266200002	&	2662	&	1.57	&	276761	\\
71	&	9169900004	&	1699	&	4.56	&	804166	&	&	226	&	9266700002	&	2667	&	3.01	&	531492	\\
72	&	9170200002	&	1702	&	2.06	&	362767	&	&	227	&	9267800004	&	2678	&	4.55	&	801427	\\
73	&	9170300002	&	1703	&	4.63	&	816310	&	&	228	&	9268600003	&	2686	&	6.46	&	1139449	\\
74	&	9170500003	&	1705	&	3.97	&	700703	&	&	229	&	9268900002	&	2689	&	5.64	&	993993	\\
75	&	9171000002	&	1710	&	5.26	&	926909	&	&	230	&	9269100003	&	2691	&	1.24	&	219322	\\
76	&	9171000003	&	1710	&	2.56	&	450836	&	&	231	&	9269300002	&	2693	&	4.37	&	769649	\\
77	&	9171100004	&	1711	&	1.28	&	225877	&	&	232	&	9270200002	&	2702	&	2.88	&	508607	\\
78	&	9171600003	&	1716	&	4.29	&	756501	&	&	233	&	9272100002	&	2721	&	6.61	&	1165708	\\
79	&	9172300002	&	1723	&	4.09	&	720996	&	&	234	&	9272200003	&	2722	&	3.91	&	689470	\\
80	&	9175700002	&	1757	&	4.69	&	826810	&	&	235	&	9272300003	&	2723	&	1.79	&	316186	\\
81	&	9176600004	&	1766	&	1.97	&	346701	&	&	236	&	9272400004	&	2724	&	6.2	&	1093794	\\
82	&	9176800004	&	1768	&	1.7	&	300522	&	&	237	&	9273200003	&	2732	&	4.78	&	842060	\\
83	&	9176900004	&	1769	&	1.62	&	285310	&	&	238	&	9273400004	&	2734	&	6.58	&	1160738	\\
84	&	9177600004	&	1776	&	3.42	&	603085	&	&	239	&	9274300003	&	2743	&	5.68	&	1002279	\\
85	&	9178100003	&	1781	&	1.75	&	309354	&	&	240	&	9276100002	&	2761	&	0.95	&	167169	\\
86	&	9179300002	&	1793	&	3.36	&	591642	&	&	241	&	9276400002	&	2764	&	2.79	&	492081	\\
87	&	9180400003	&	1804	&	2.58	&	454489	&	&	242	&	9276600002	&	2766	&	3.79	&	668182	\\
88	&	9180700003	&	1807	&	4.22	&	744111	&	&	243	&	9276600003	&	2766	&	3.47	&	611061	\\
89	&	9181400002	&	1814	&	1.19	&	209206	&	&	244	&	9276700003	&	2767	&	4.7	&	828186	\\
90	&	9181700003	&	1817	&	2.36	&	415451	&	&	245	&	9278000004	&	2780	&	3.44	&	607388	\\
91	&	9181900003	&	1819	&	4.2	&	740381	&	&	246	&	9278900002	&	2789	&	3.99	&	703211	\\
92	&	9182100003	&	1821	&	5.27	&	928537	&	&	247	&	9279400003	&	2794	&	0.94	&	166137	\\
93	&	9182200003	&	1822	&	4.81	&	848729	&	&	248	&	9280600003	&	2806	&	1.98	&	349925	\\
94	&	9182500003	&	1825	&	3.03	&	534405	&	&	249	&	9281000002	&	2810	&	3.13	&	552697	\\
95	&	9185700003	&	1857	&	1.52	&	267126	&	&	250	&	9281200003	&	2812	&	3.62	&	637549	\\
96	&	9187200003	&	1872	&	2.3	&	405573	&	&	251	&	9281300003	&	2813	&	2.18	&	384111	\\
97	&	9187300003	&	1873	&	3.35	&	591401	&	&	252	&	9285000003	&	2850	&	5.78	&	1018841	\\
98	&	9187400002	&	1874	&	4.27	&	751985	&	&	253	&	9285400002	&	2854	&	1	&	176060	\\
99	&	9187400003	&	1874	&	2.52	&	443917	&	&	254	&	9285400003	&	2854	&	3.17	&	559730	\\
100	&	9188300003	&	1883	&	1.09	&	192227	&	&	255	&	9285600002	&	2856	&	0.98	&	173367	\\
101	&	9189000003	&	1890	&	2.53	&	446735	&	&	256	&	9285700003	&	2857	&	3.8	&	669560	\\
102	&	9189200004	&	1892	&	3.79	&	668616	&	&	257	&	9288200003	&	2882	&	1.08	&	191254	\\
103	&	9190100002	&	1901	&	5.01	&	882669	&	&	258	&	9289500004	&	2895	&	1.73	&	305505	\\
104	&	9190400003	&	1904	&	1.39	&	244955	&	&	259	&	9289800002	&	2898	&	1.13	&	200067	\\
105	&	9190600003	&	1906	&	2.32	&	409140	&	&	260	&	9290800002	&	2908	&	3.45	&	608332	\\
106	&	9191000002	&	1910	&	6.44	&	1134925	&	&	261	&	9291100003	&	2911	&	3	&	529685	\\
107	&	9191100005	&	1911	&	3.12	&	549656	&	&	262	&	9291500002	&	2915	&	1.03	&	181709	\\
108	&	9191300004	&	1913	&	4.98	&	877200	&	&	263	&	9291600004	&	2916	&	3.01	&	530927	\\
109	&	9191600003	&	1916	&	2.83	&	498338	&	&	264	&	9291800002	&	2918	&	4.92	&	867689	\\
110	&	9191700004	&	1917	&	5.73	&	1010187	&	&	265	&	9291900002	&	2919	&	2.67	&	470591	\\
111	&	9191800002	&	1918	&	5.73	&	1010697	&	&	266	&	9292200002	&	2922	&	5.23	&	921950	\\
112	&	9192100003	&	1921	&	2.25	&	397558	&	&	267	&	9292300002	&	2923	&	3.36	&	592678	\\
113	&	9193100002	&	1931	&	2.14	&	377032	&	&	268	&	9292300003	&	2923	&	3.32	&	585432	\\
114	&	9193200002	&	1932	&	3.09	&	545221	&	&	269	&	9292400005	&	2924	&	3.6	&	635465	\\
115	&	9194500007	&	1945	&	2.16	&	380834	&	&	270	&	9293100002	&	2931	&	6.35	&	1118778	\\
116	&	9194800004	&	1948	&	3.55	&	625570	&	&	271	&	9293400002	&	2934	&	1.11	&	196290	\\
117	&	9195000003	&	1950	&	3.73	&	658118	&	&	272	&	9293500002	&	2935	&	1.21	&	213551	\\
118	&	9196600002	&	1966	&	5.18	&	913961	&	&	273	&	9293700002	&	2937	&	4.79	&	844386	\\
119	&	9196900002	&	1969	&	1.25	&	220960	&	&	274	&	9294700014	&	2947	&	1.85	&	325342	\\
120	&	9197000002	&	1970	&	5.27	&	929152	&	&	275	&	9294800004	&	2948	&	1.6	&	282729	\\
121	&	9197500003	&	1975	&	2.27	&	399390	&	&	276	&	9294900005	&	2949	&	4.03	&	710343	\\
122	&	9198100002	&	1981	&	2.22	&	390596	&	&	277	&	9305600003	&	3056	&	4.35	&	767630	\\
123	&	9198300002	&	1983	&	4.06	&	716621	&	&	278	&	9305600004	&	3056	&	4.58	&	807780	\\
124	&	9198400003	&	1984	&	2.51	&	443394	&	&	279	&	9305700003	&	3057	&	5.35	&	943508	\\
125	&	9198700006	&	1987	&	1.09	&	192494	&	&	280	&	9305700005	&	3057	&	2	&	352481	\\
126	&	9198900002	&	1989	&	1.82	&	320276	&	&	281	&	9305800002	&	3058	&	4.78	&	843571	\\
127	&	9198900004	&	1989	&	4.43	&	780710	&	&	282	&	9306300003	&	3063	&	2.3	&	405608	\\
128	&	9199200004	&	1992	&	1.05	&	185637	&	&	283	&	9306400004	&	3064	&	6.67	&	1176203	\\
129	&	9199500004	&	1995	&	1.91	&	336064	&	&	284	&	9307500002	&	3075	&	2.49	&	438811	\\
130	&	9200100005	&	2001	&	3.19	&	563143	&	&	285	&	9307800002	&	3078	&	1.62	&	285357	\\
131	&	9200200002	&	2002	&	3.23	&	568866	&	&	286	&	9307900004	&	3079	&	3.1	&	546744	\\
132	&	9200400003	&	2004	&	2.12	&	373047	&	&	287	&	9307900005	&	3079	&	6.48	&	1142453	\\
133	&	9200800004	&	2008	&	4.74	&	835613	&	&	288	&	9308100004	&	3081	&	3.64	&	641870	\\
134	&	9200900003	&	2009	&	1.74	&	307235	&	&	289	&	9308100005	&	3081	&	2.51	&	442592	\\
135	&	9201300003	&	2013	&	4.14	&	730250	&	&	290	&	9308700003	&	3087	&	3.16	&	557727	\\
136	&	9201400003	&	2014	&	1.66	&	292367	&	&	291	&	9309200003	&	3092	&	2.53	&	446555	\\
137	&	9201500003	&	2015	&	6.49	&	1144810	&	&	292	&	9309900002	&	3099	&	4.56	&	804260	\\
138	&	9202100003	&	2021	&	1.87	&	330235	&	&	293	&	9310100003	&	3101	&	7.66	&	1349994	\\
139	&	9202900002	&	2029	&	4.21	&	741632	&	&	294	&	9310200004	&	3102	&	4.71	&	830686	\\
140	&	9204700002	&	2047	&	1.6	&	282537	&	&	295	&	9311100002	&	3111	&	1.56	&	274181	\\
141	&	9205700003	&	2057	&	6.36	&	1121529	&	&	296	&	9311100005	&	3111	&	5.01	&	883409	\\
142	&	9207100003	&	2071	&	2.63	&	463072	&	&	297	&	9312000002	&	3120	&	0.97	&	170493	\\
143	&	9207600004	&	2076	&	1.7	&	299695	&	&	298	&	9312000003	&	3120	&	3.08	&	542823	\\
144	&	9207700003	&	2077	&	3.48	&	612781	&	&	299	&	9312000004	&	3120	&	2.03	&	358103	\\
145	&	9208100004	&	2081	&	5.96	&	1051690	&	&	300	&	9313500002	&	3135	&	4.54	&	800551	\\
146	&	9208400003	&	2084	&	5.35	&	942755	&	&	301	&	9313900002	&	3139	&	2.8	&	492855	\\
147	&	9209500004	&	2095	&	3.19	&	561571	&	&	302	&	9315100002	&	3151	&	1.52	&	267559	\\
148	&	9209600002	&	2096	&	6.56	&	1156232	&	&	303	&	9316200002	&	3162	&	4.86	&	857242	\\
149	&	9209800002	&	2098	&	5.43	&	957569	&	&	304	&	9316200003	&	3162	&	1.55	&	273718	\\
150	&	9210100003	&	2101	&	4.22	&	744294	&	&	305	&	9317200002	&	3172	&	1.54	&	271656	\\
151	&	9210700004	&	2107	&	3.95	&	696954	&	&	306	&	9319100004	&	3191	&	4.31	&	760519	\\
152	&	9211000002	&	2110	&	4.87	&	857875	&	&	307	&	9321200004	&	3212	&	3.71	&	654218	\\
153	&	9211600002	&	2116	&	5.94	&	1047642	&	&	308	&	9321700003	&	3217	&	1.98	&	348438	\\
154	&	9211700002	&	2117	&	2.14	&	376952	&	&										
\enddata
\vspace{4mm}
\end{deluxetable*}
\begin{table}
\centering
\caption{Pointed \xmm\ EPIC-pn  Small Window Mode Observations}
\label{table:obs_pointed}
\begin{tabular}{lccccc}
\hline\hline
Source & Obs. ID  & Year & Filter & Exp. & Number of \\
    &       & & Set up & ks & Frames ($\times 10^{6}$)\\
    \hline
AB~Doradus & 0134522101  & 2002 & Closed Flt. & 49 & 8.55\\
AB~Doradus & 0160362901  & 2004 & Closed Flt. & 56 & 9.87\\
AB~Doradus & 0791980401  & 2017 & Thick Flt. & 12 & 2.08\\
SNR 21.5$-$09 & 0804250201 & 2017 & Thin Flt. & 41 & 7.14 \\
\hline\hline

\end{tabular}
\end{table}

\software{%
 \texttt{GEANT4} \citep{tenzer2010},
 \texttt{\xmm\ SAS} \citep{gabriel2004},
 \texttt{Matplotlib} \citep{Hunter:2007},
 \texttt{NumPy} \citep{van2011numpy},
 \texttt{Astropy} \citep{2013A&A...558A..33A},
 \texttt{SciPy} \citep{jones_scipy_2001}.
}

\bibliographystyle{aasjournal}
\bibliography{literature}

\end{document}